\documentclass[twocolumn]{aastex62}
\usepackage{amsmath}     
\usepackage{wasysym}           
\usepackage{graphicx}
\usepackage{amssymb}
\usepackage{epstopdf}
\usepackage{mathrsfs}
\usepackage{anyfontsize}
\usepackage{natbib}
\usepackage{color}
\usepackage{lipsum}
\usepackage{diagbox}

\shorttitle{Stars Crushed by Black Holes II}
\shortauthors{Coughlin \& Nixon}
\begin{document}
\title{Stars Crushed by Black Holes.~II.~A Physical Model of Adiabatic Compression and Shock Formation in Tidal Disruption Events}
\author[0000-0003-3765-6401]{E.~R.~Coughlin}
\affiliation{Department of Physics, Syracuse University, Syracuse, NY 13244}
\author[0000-0002-2137-4146]{C.~J.~Nixon}
\affiliation{Department of Physics and Astronomy, University of Leicester, Leicester, LE1 7RH, UK}

\email{ecoughli@syr.edu}

\begin{abstract}
We develop a Newtonian model of a deep tidal disruption event (TDE), for which the pericenter distance of the star, $r_{\rm p}$, is well within the tidal radius of the black hole, $r_{\rm t}$, i.e., when $\beta \equiv r_{\rm t}/r_{\rm p} \gg 1$. We find that shocks form for $\beta \gtrsim 3$, but they are weak (with Mach numbers $\sim 1$) for all $\beta$, and that they reach the center of the star prior to the time of maximum adiabatic compression for $\beta \gtrsim 10$. The maximum density and temperature reached during the TDE follow much shallower relations with $\beta$ than the previously predicted $\rho_{\rm max} \propto \beta^3$ and $T_{\rm max} \propto \beta^2$ scalings. Below $\beta \simeq 10$, this shallower dependence occurs because the pressure gradient is dynamically significant before the pressure is comparable to the ram pressure of the freefalling gas, while above $\beta \simeq 10$ we find that shocks prematurely halt the compression and yield the scalings $\rho_{\rm max} \propto \beta^{1.62}$ and $T_{\rm max} \propto \beta^{1.12}$. We find excellent agreement between our results and high-resolution simulations. Our results demonstrate that, in the Newtonian limit, the compression experienced by the star is completely independent of the mass of the black hole. We discuss our results in the context of existing (affine) models, polytropic vs.~non-polytropic stars, and general relativistic effects, which become important when the pericenter of the star nears the direct capture radius, at $\beta \sim 12.5$ (2.7) for a solar-like star disrupted by a $10^6M_{\odot}$ ($10^{7}M_{\odot}$) supermassive black hole. 
\end{abstract}

\keywords{Astrophysical black holes (98) --- Black hole physics (159) --- Hydrodynamics (1963) --- Shocks (2086) --- Supermassive black holes (1663) --- Tidal disruption (1696)}

\section{Introduction}
High-cadence, wide-field surveys are discovering tidal disruption events (TDEs) -- where a star is destroyed by the tidal field of a supermassive black hole and the accretion of the stellar debris briefly illuminates the host galaxy -- at an extremely accelerated rate.  Indeed, the number of definitive detections now exceeds 50, and observations continue to yield a wealth of data about the physical processes at play during these events (e.g., \citealt{Bade:1996aa, Komossa:1999aa, Esquej:2007aa, Gezari:2009aa, Gezari:2012aa, Holoien:2014aa, Miller:2015ab, Vinko:2015aa, Alexander:2016aa, Cenko:2016aa, Holoien:2016aa, jiang16, Kara:2016aa, vanVelzen:2016aa, Alexander:2017aa, Blanchard:2017aa, Brown:2017aa, Gezari:2017aa, Hung:2017aa, Saxton:2017aa, Brown:2018aa, Pasham:2018aa, Blagorodnova:2019aa, Hung:2019aa, Holoien:2019aa, Leloudas:2019aa, Nicholl:2019aa, Pasham:2019aa, Saxton:2019aa, Hung:2020aa, Hung:2020ab, Holoien:2020aa, Jonker:2020aa, Kajava:2020aa, Li:2020aa, Hinkle:2021aa, jiang21, Payne:2021aa, vanVelzen:2021aa}); see also the recent reviews by \citet{Alexander:2020aa, vanVelzen:2020aa, Saxton:2020aa, Gezari:2021aa} and references therein.

The outcome of the tidal interaction between any two objects is largely controlled by their point of closest approach. In the TDE literature this distance is usually defined implicitly via $\beta \equiv r_{\rm t}/r_{\rm p}$, where $r_{\rm t} = R_{\star}\left(M_{\bullet}/M_{\star}\right)^{1/3}$ is $\sim$ the separation between the two bodies at which the self-gravity of the star equals the tidal force of the hole, and $r_{\rm p}$ is the point of closest approach between the stellar center of mass and the black hole. Because the tidal force decays as the inverse cube of the distance between the star and the supermassive black hole, changes in $\beta$ by a factor of the order unity can dramatically alter the end state of a TDE. 

When $\beta \lesssim 1$, tides may be insufficient to completely unbind the star, resulting in a partial TDE. \citet{Guillochon:2013aa} (see also \citealt{Mainetti:2017aa}) showed that, for a $\Gamma = \gamma = 5/3$ polytrope, TDEs with $\beta \lesssim 0.9$ resulted in the partial disruption of the star, while $\beta \gtrsim 2$ was necessary to completely destroy a $\gamma = 5/3$, $\Gamma = 4/3$ polytrope (we denote the polytropic index by $\Gamma$, such that $p \propto \rho^{\Gamma}$ with $p$ the pressure and $\rho$ the density, while we let $\gamma$ be the adiabatic index; see \citealt{Golightly:2019ab, Law-Smith:2020aa, Nixon:2021ab} for how the critical $\beta$ for full disruption varies with more realistic stellar profiles). The surviving core has a profound effect on the return of disrupted debris to the supermassive black hole \citep{Guillochon:2013aa}, and generates a power-law decline of the fallback rate that is $\propto t^{-9/4}$ \citep{Coughlin:2019aa, Miles:2020aa, Nixon:2021ab} -- distinct from the canonical $\propto t^{-5/3}$ scaling \citep{Rees:1988aa, Phinney:1989aa}. 

On the other hand, a TDE with $\beta \gg 1$ implies that the gravitational field of the supermassive black hole completely overwhelms the self-gravity of the star. When the star enters the tidal sphere of the black hole, the tidal force compresses the star in the direction perpendicular to the star's orbital plane, stretch the star approximately in the direction connecting the stellar center of mass and the supermassive black hole, and compress the star in the direction within the plane and orthogonal to that line. The latter two effects roughly compensate for one another, meaning that the compression out of the plane is largely responsible for increasing the central density of the star during its descent. \cite{carter83} argued that the increasing temperature and density could ignite thermonuclear runaway if the tidal encounter is sufficiently deep; by assuming that the gas evolves adiabatically during the compression and equating the ram pressure of the infalling gas to the thermal pressure at the point of maximum compression, they derived that the maximum density achieved by tidally squeezing the star is $\rho_{\rm max} \propto \beta^{2/(\gamma-1)}$. \citet{Bicknell:1983aa} rebuked this idea and argued that the crushing of the stellar envelope by the tidal force would be accompanied by the formation of a shock that would prematurely halt the adiabatic compression, resulting in a much less dramatic increase in the central density and temperature of the star. Their smoothed-particle hydrodynamics (SPH) simulations seemed to validate their argument, and showed a much weaker dependence of the maximum-achieved density on $\beta$, with $\rho_{\rm max} \propto \beta^{1.5}$ fitting the results of their simulations up to $\beta \sim 10$. Above $\beta \sim 10$ their simulations showed a declining maximum density with $\beta$. 

There have since been a number of other investigations into the nature of the increase in the central density during a deep tidal encounter. \citet{laguna93} noted that their SPH simulations yielded an exponent relating the maximum central density to $\beta$ that was between 1.5 and 2, and thus significantly weaker than the $\propto \beta^3$ scaling (for $\gamma = 5/3$) predicted by \citet{carter83}. \citet{brassart08} performed one-dimensional, hydrodynamic simulations of disruptions of polytropic stars up to $\beta = 15$ and found good agreement with the predicted $\rho \propto \beta^3$ scaling for $\gamma = 5/3$. \citet{Stone:2013aa}, using an analytical model, found the $\rho_{\rm max} \propto \beta^{2/(\gamma-1)}$ scaling initially derived by \citet{carter83} (and also derived in Section II of \citealt{Bicknell:1983aa}). {}{Most recently, \citet{gafton19} investigated non-relativistic and relativistic disruptions of a solar-like, $\gamma=5/3$ polytrope by a $10^{6}M_{\odot}$ black hole; they found $\rho_{\rm max}/\rho_{\rm c} \propto \beta^{1.7}$ and $\rho_{\rm max}/\rho_{\rm c} \propto \beta^{1.85}$ for $\beta \lesssim 4$ in the Newtonian and general relativistic regimes, respectively, while they found much shallower increases above $\beta = 4$ ($\propto \beta^{0.65}$ in the Newtonian limit, and $\propto \beta ^{0.2}$, $\propto \beta^{0.5}$, and $\propto \beta^{1}$ for retrograde-Kerr, Schwarzschild, and prograde-Kerr, respectively).} Thus, there does not yet appear to be a consensus on the maximum central density and temperature that the star can attain during a deeply penetrating TDE or, indeed, whether the thermonuclear ignition originally envisaged by \citet{carter83} can occur. 

Here we develop a hydrodynamical model for the evolution of the star during a TDE. In Section \ref{sec:pressureless} we briefly consider the case where the star is treated as a collection of freefalling, non-interacting particles during the approach of the star to pericenter in a deep TDE; doing so allows us to establish some basic relationships that we use to motivate the work in later sections. We also demonstrate that the {pressure gradient} within the star -- which is responsible for reversing the tidal compression -- becomes important much earlier in the tidal encounter than the pressure itself, which leads to a weaker dependence of the maximum-achievable density on $\beta$ compared to that recovered by equating the gas pressure to the ram pressure. 

In Section \ref{sec:homologous} we demonstrate that a homologous relationship between the current and initial positions of a Lagrangian fluid element, while exactly satisfied in the pressureless-freefall limit, is upheld to leading order in $z_0$ (the initial height of a fluid element) when the pressure and self-gravity of the collapsing star are included in the equation of motion for the fluid elements. This homologous relationship yields a much flatter dependence of $\rho_{\rm max}$ on $\beta$ than $\rho \propto \beta^{3}$ up until $\beta \simeq 10$, beyond which the scaling $\rho_{\rm max} \propto \beta^3$ is recovered, but with a proportionality constant that is a factor of $\sim 5$ smaller than the one predicted by \citet{luminet86}.

In Section \ref{sec:non-homologous} we develop and solve the equations that describe the nonlinear (non-homologous) relationship between the current and initial positions of a fluid element, and in Section \ref{sec:shocks} we show that this nonlinearity leads to the formation of shocks within the flow when $\beta \gtrsim 3$ for a $\gamma = 5/3$ polytrope. These shocks serve to significantly weaken the $\beta$-dependence of $\rho_{\rm max}$ and $T_{\rm max}$, the maximum density and temperature achieved by the compressing star, for $\beta \gtrsim 10$. Consequently, the $\rho_{\rm max} \propto \beta^3$ and $T_{\rm max} \propto \beta^2$ scalings are not realized during the deep tidal encounter of a star with a supermassive black hole. We also compare our findings to SPH simulations (presented in full in \citealt{norman21}) and find excellent agreement with our predictions. 

In Section \ref{sec:discussion} we compare and contrast our model and approach to the affine-star prescriptions that have been used in previous works, including that of \citet{carter83} who used such an affine-star approximation to study deep TDEs. We also discuss the consequences of relaxing the polytropic assumption that we make in developing the model in Sections \ref{sec:homologous} -- \ref{sec:shocks}, and we show the results for the homologous compression of a $\gamma = 5/3$ and $\Gamma = 4/3$ polytrope. 

In this paper and in the models that we develop here, we use a purely Newtonian treatment: the self-gravity of the star is treated non-relativistically (such that the self-gravitational potential satisfies the usual Poisson equation and the gravitational force is the gradient of the potential) and the gravitational field of the black hole is modeled with the potential of a Newtonian point mass. The reasons for doing so are for ease of comparison with previous works (one of our goals is to resolve the previously noted, conflicting claims about the compression of the star in the high-$\beta$ limit where the analyses have been made in the Newtonian approximation), to make comparisons with our simulations, and for simplicity (as the model we develop here is itself novel, and it is reasonable to start in the Newtonian regime before extending the work to incorporate general relativity). However, we emphasize that for typical stars (i.e., for low-mass stars with mass and radii on the order of or less than solar values) disrupted by high-mass black holes ($M_{\bullet} \gtrsim 10^{6}M_{\odot}$), general relativistic effects can be important even for modest $\beta$ \citep{gafton19}, and large values of $\beta$ can result in the direct capture of the star or its plummet into the event horizon (where it is still, presumably, destroyed by tides, but with zero observational relevance). We derive an order-of-magnitude estimate of the discrepancy between the Newtonian and general relativistic tides, the range of parameters within which we expect the star to be directly captured by the black hole, and we compare to previous works in Section \ref{sec:gr}.

We summarize and conclude in Section \ref{sec:summary}.


\section{Pressureless freefall solutions}
\label{sec:pressureless}
Here we analyze the vertical compression of the star in the limit that the star is a collection of freely falling particles in the tidal field of the hole. We adopt the tidal approximation such that the motion of the center of mass of the star decouples from the motion of its extremities. The equation for the distance of the center of mass of the star from the black hole, which we denote $r_{\rm c}$, is then (making the usual assumption that the star is on a parabolic orbit)

\begin{equation}
\frac{1}{2}\left(\frac{\partial r_{\rm c}}{\partial t}\right)^{2}+\frac{GM_{\bullet}r_{\rm p}}{r_{\rm c}^2}-\frac{GM_{\bullet}}{r_{\rm c}} = 0. \label{rceq}
\end{equation}
The solution to this equation is 

\begin{equation}
r_{\rm c} = r_{\rm p}\cosh^2\left(\tau\right), \label{uceq}
\end{equation}
where $\tau$ is defined as 

\begin{equation}
\frac{\partial \tau}{\partial t} = \sqrt{\frac{GM_{\bullet}}{2r_{\rm c}^{3}}}.
\end{equation}
and $r_{\rm p} \equiv r_{\rm t}/\beta$ is the pericenter distance of the star. The equation of motion for gas parcels out of the plane is 

\begin{equation}
\frac{\partial^2z}{\partial t^2} = -\frac{GM_{\bullet}}{r_{\rm c}(t)^3}z,
\end{equation}
and changing variables to $\tau$ turns this into

\begin{equation}
\ddot{z}-3\tanh(\tau)\dot{z}+2z = 0, \label{zeq}
\end{equation}
where dots denote differentiation with respect to $\tau$. If we assume that the star retains perfect hydrostatic balance until reaching the tidal radius \citep{lacy82, Stone:2013aa}, then $z(\tau_{\rm t}) = z_0$ and $\dot{z}(\tau_{\rm t}) = 0$, where $z_0$ is the initial height of a fluid element out of the orbital plane and $\tau_{\rm t} = -{\rm arcsinh}\left(\sqrt{\beta-1}\right)$ is the time at which the center of mass is at the tidal radius. The solution to Equation \eqref{zeq} that satisfies these initial conditions is

\begin{equation}
\frac{z}{z_0} = \frac{1-2\sqrt{\beta-1}\sinh(\tau)-\sinh^2\left(\tau\right)}{\beta} \equiv H(\tau). \label{Heq}
\end{equation}

By solving the $x$ and $y$ equations of motion we can recover the density. However, as pointed out by \citet{carter83}, the motion within the plane is approximately area-conserving, and hence the motion out of the plane contributes predominantly to changes in the density. We can therefore approximate the motion of the collapsing fluid as one-dimensional and perpendicular to the orbital plane. The integrated mass to any position $z$ above the plane is then a conserved Lagrangian quantity, and hence the density is

\begin{equation}
\rho(z,\tau) = \frac{\partial z_0}{\partial z}\rho_0\left[z_0 = z_0(z,\tau)\right] = \frac{1}{H(\tau)}\rho_0\left[{z}/{H(\tau)}\right],
\end{equation}
where $\rho_0$ is the density profile of the original star. If the progenitor is spherically symmetric with a well-behaved maximum in its density at its geometric center we can write, without loss of generality, 

\begin{equation}
\rho_0(z_0) \simeq \rho_{\rm c}\left(1-\frac{s_0^2+z_0^2}{\alpha^2}+\mathcal{O}\left[\left(s_0^2+z_0^2\right)^2\right]\right), 
\end{equation}
where $s_0$ is the cylindrical radius from the vertical axis and $\alpha$ is a star-specific scale length; for a polytrope it follows from hydrostatic balance that

\begin{equation}
\alpha^2 = \frac{6\gamma p_{\rm c}}{\rho_{\rm c}}\frac{1}{4\pi G\rho_{\rm c}}, \label{alphaeq}
\end{equation} 
where $p_{\rm c}$ is the central pressure. Since {we never specified a length scale for $z_0$} we can simply let $z_0 \rightarrow \alpha z_0$. With $s_0 \simeq s$ the density directly above the center of the star (with $s_0 = 0$) is

\begin{equation}
\rho(z,\tau) = \frac{\rho_{\rm c}}{H(\tau)}\left(1-\frac{z^2}{H(\tau)^2}\right). \label{rhoz}
\end{equation}
For adiabatic compression the pressure satisfies

\begin{equation}
p = S_0(z_0)\rho^{\gamma},
\end{equation}
where $S_0(z_0)$ is the stellar entropy profile, and for a constant $S_0$ the pressure to leading order is

\begin{equation}
p = p_{\rm c} H^{-\gamma}\left(1-\gamma\frac{z^2}{H^2}\right). \label{pz}
\end{equation}

We can now estimate when the pressure becomes important in modifying the dynamical evolution of the collapsing star. \citet{carter83} asserted that the pressure is important when it becomes comparable to the ram pressure of the compressing gas; one can use Equation \eqref{Heq} to determine the velocity, and with Equations \eqref{rhoz} and \eqref{pz} for the density and pressure we recover their solution, namely that $H_{\rm eq} \propto \beta^{-2/(\gamma-1)}$ at which this equality occurs. However, it is not the pressure itself that resists the compression but the pressure gradient, and the pressure gradient is significant when

\begin{equation}
\frac{1}{\rho}\frac{\partial p}{\partial z} \simeq -\frac{GM_{\bullet}z}{r_{\rm c}^3}.
\end{equation}
If we use the previously derived expressions for the density and pressure, then carrying out the algebraic manipulations and assuming that the equality occurs near pericenter, such that $r_{\rm c} \simeq r_{\rm p} = r_{\rm t}/\beta$, we find that the pressure gradient equals the tidal force when

\begin{equation}
H_{\rm eq} \propto \beta^{-\frac{3}{\gamma+1}} \quad \Rightarrow \quad \rho_{\rm eq} \propto \beta^{\frac{3}{\gamma+1}}.
\end{equation}
This scaling is {much shallower} than $\beta^{2/(\gamma-1)}$; for example, with $\gamma = 5/3$ we find $H_{\rm eq} \propto \beta^{-9/8}$. This weaker dependence occurs because the compression of the star leads to a much more rapid increase in the pressure gradient compared to the pressure, i.e., $\partial p/\partial z \simeq p_{\rm c}/H$.

Various aspects of the pressureless collapse have been delimited by \citet{Stone:2013aa} (see also Section II of \citealt{Bicknell:1983aa}); we refer the interested reader to their work on the subject and move on to the inclusion of pressure and self-gravity. As we show below, these effects significantly impact the simple scaling relations derived above.

\section{Pressurized and self-gravitating, homologous solutions}
\label{sec:homologous}
The freefall solutions of the previous section demonstrate that the Lagrangian height of a fluid element is related to its initial height by

\begin{equation}
z = H(\tau)z_0, \label{zhomologous}
\end{equation}
and hence the evolution of the compressing star is homologous. While we derived this relationship in the freefall case, this relationship must hold {to leading order in $z_0$}, i.e., near the midplane, with pressure and self-gravity modifying the motion of the gas. With Equation \eqref{zhomologous} the density is still given by

\begin{equation}
\rho = \rho_{\rm c}\frac{\partial z_0}{\partial z}\left(1-\left({s_0^2+z_0^2}\right)\right) \equiv \frac{\rho_{\rm c}}{H(\tau)}g(s_0,z_0).
\end{equation}
The gravitational potential $\Phi$ solves:

\begin{equation}
\nabla^2\Phi = 4\pi G\rho.
\end{equation}
To leading order in this equation $\rho$ is $\rho_{\rm c}/H$, and hence the gravitational potential is simply\footnote{This method of Taylor expanding the solution to the gravitational potential about the origin ignores aspherical contributions from gas at large radii within the star. While this approximation is well upheld in the limit that the $\beta$ of the encounter is large and the star is within the tidal sphere, where the pressure of the gas dominates self-gravity (see the relative scaling of the terms in parentheses on the left-hand side of Equation \ref{Hdyn}), and/or when the star is approximately spherically symmetric, it is less accurate when the $\beta$ of the encounter is only modest. We plan to explore the modest-$\beta$ and partial-disruption limit in a future investigation.} 

\begin{equation}
\Phi = 4\pi G\rho_{\rm c}\alpha^2H^{-1}\frac{r^2}{6} \label{phisimp}, 
\end{equation}
while the pressure is

\begin{equation}
p = p_{\rm c}\left(\frac{\partial z_0}{\partial z}\right)^{\gamma}g^{\gamma} \simeq p_{\rm c}H^{-\gamma}\left(1-\gamma\left({z_0^2+s_0^2}\right)\right).
\end{equation}
Inserting these relationships into the $z$-component of the momentum equation,

\begin{equation}
\frac{\partial^2z}{\partial t^2}+\frac{1}{\rho}\frac{\partial p}{\partial z} = -\frac{GM_{\bullet}z}{r_{\rm c}^3}-\frac{\partial \Phi}{\partial z}, \label{zmom}
\end{equation}
and equating linear-order terms in $z_0$ shows that $H$ must satisfy\footnote{\label{footnote:2}Interestingly, the black hole mass does not enter into this equation, and the only dependencies are on the ratios of the central to the average density and the pericenter distance to the tidal radius. The entire tidal interaction is therefore constrained by only two dimensionless numbers (and the adiabatic index).}

\begin{equation}
\mathcal{L}\left[H\right]
-\frac{2}{\beta^3}\frac{\rho_{\rm c}}{\rho_{\star}}\left(H^{-\gamma}-1\right)\cosh^6(\tau) = 0.  \label{Hdyn}
\end{equation}
Here $\rho_{\star} = M_{\star}/(4\pi R_{\star}^3/3)$ is the average stellar density and

\begin{equation}
\mathcal{L} = \frac{\partial^2}{\partial \tau^2}-3\tanh(\tau)\frac{\partial}{\partial \tau}+2.
\end{equation}
This equation includes the effects of both pressure and self-gravity, and therefore (instead of using the impulse approximation) the initial condition is that the star is in hydrostatic balance infinitely far from the black hole (as $r_{\rm c} \rightarrow \infty$), so $H(\tau \rightarrow -\infty) = 1$ and $\dot{H}(\tau\rightarrow-\infty) = 0$.  

\begin{figure*}[htbp]
\centering
\includegraphics[width=0.495\textwidth]{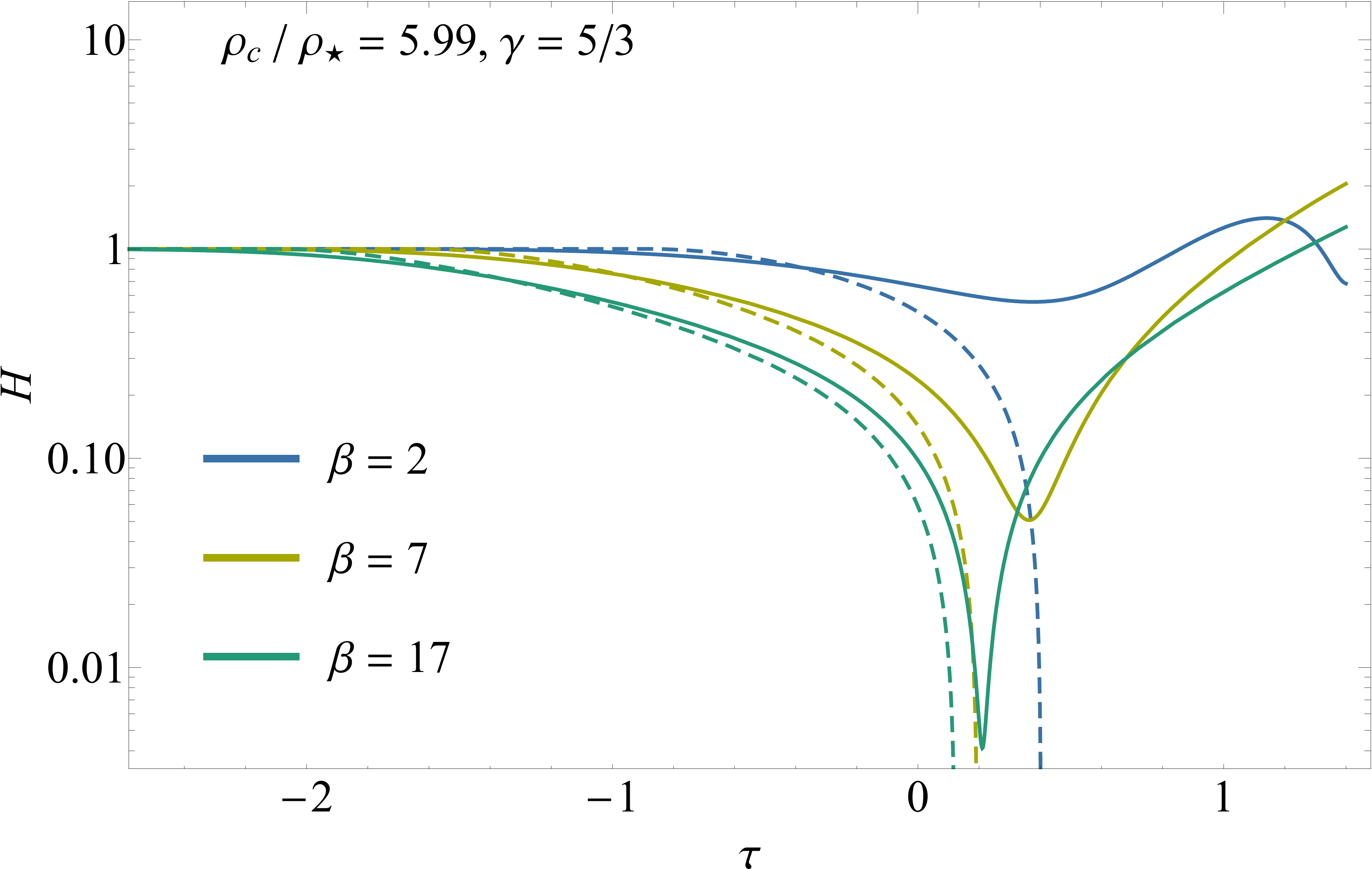}
\includegraphics[width=0.495\textwidth]{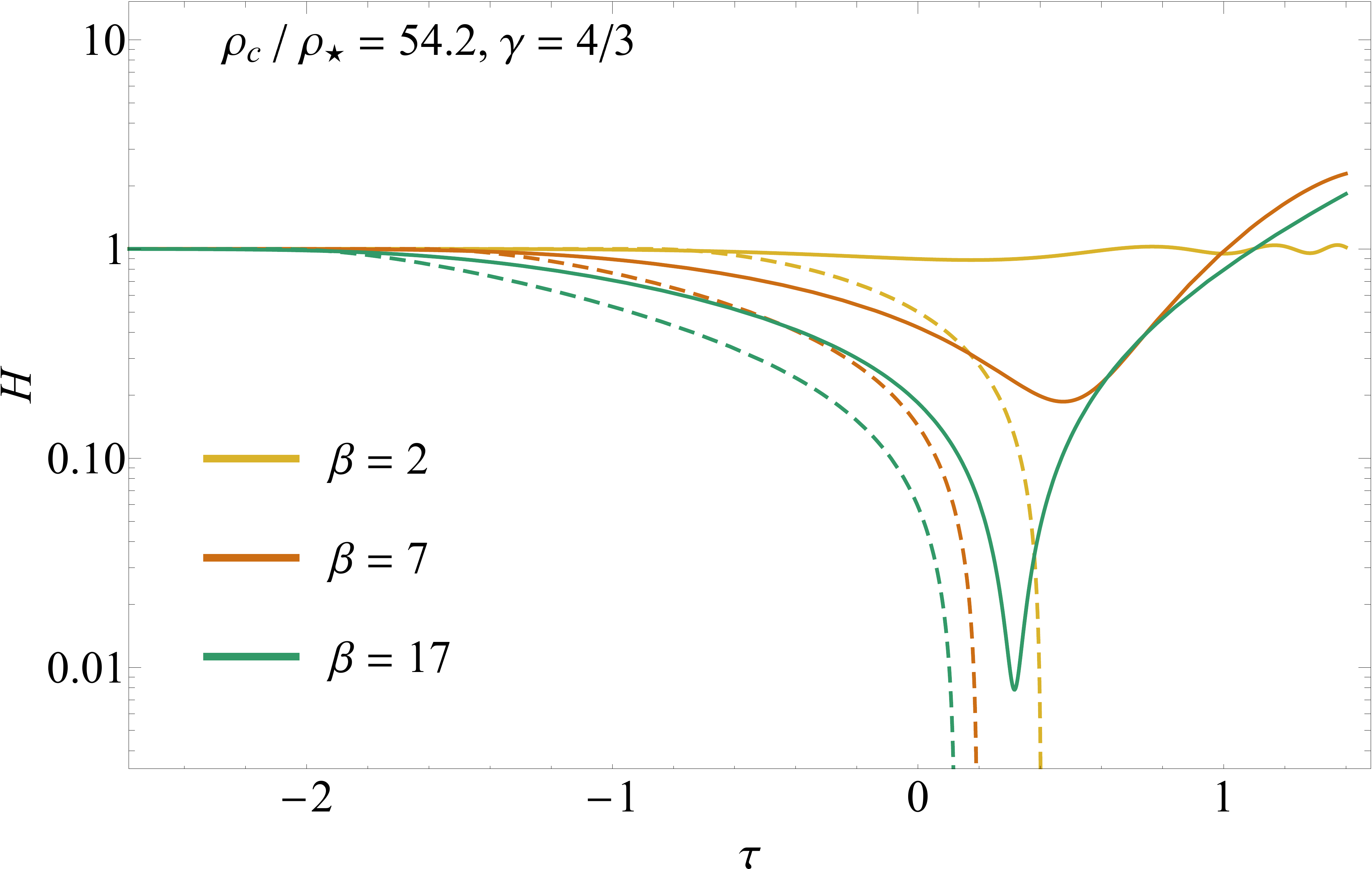}
\caption{The solution for $H$ that results from the assumption that the gas freefalls in the tidal field of the black hole (dashed curves) and the solution that incorporates gas pressure and self-gravity (solid curves) for a $\gamma = 5/3$ polytrope (left) and a $\gamma = 4/3$ polytrope (right). This shows that the pressure of the gas is important in resisting compression for times significantly earlier than when the star would be compressed to zero height ($H = 0$) in the absence of pressure; this time coincides with when the dashed lines are nearly vertical.}
\label{fig:H_Hff}
\end{figure*}

\begin{figure*}[htbp] 
   \centering
   \includegraphics[width=0.495\textwidth]{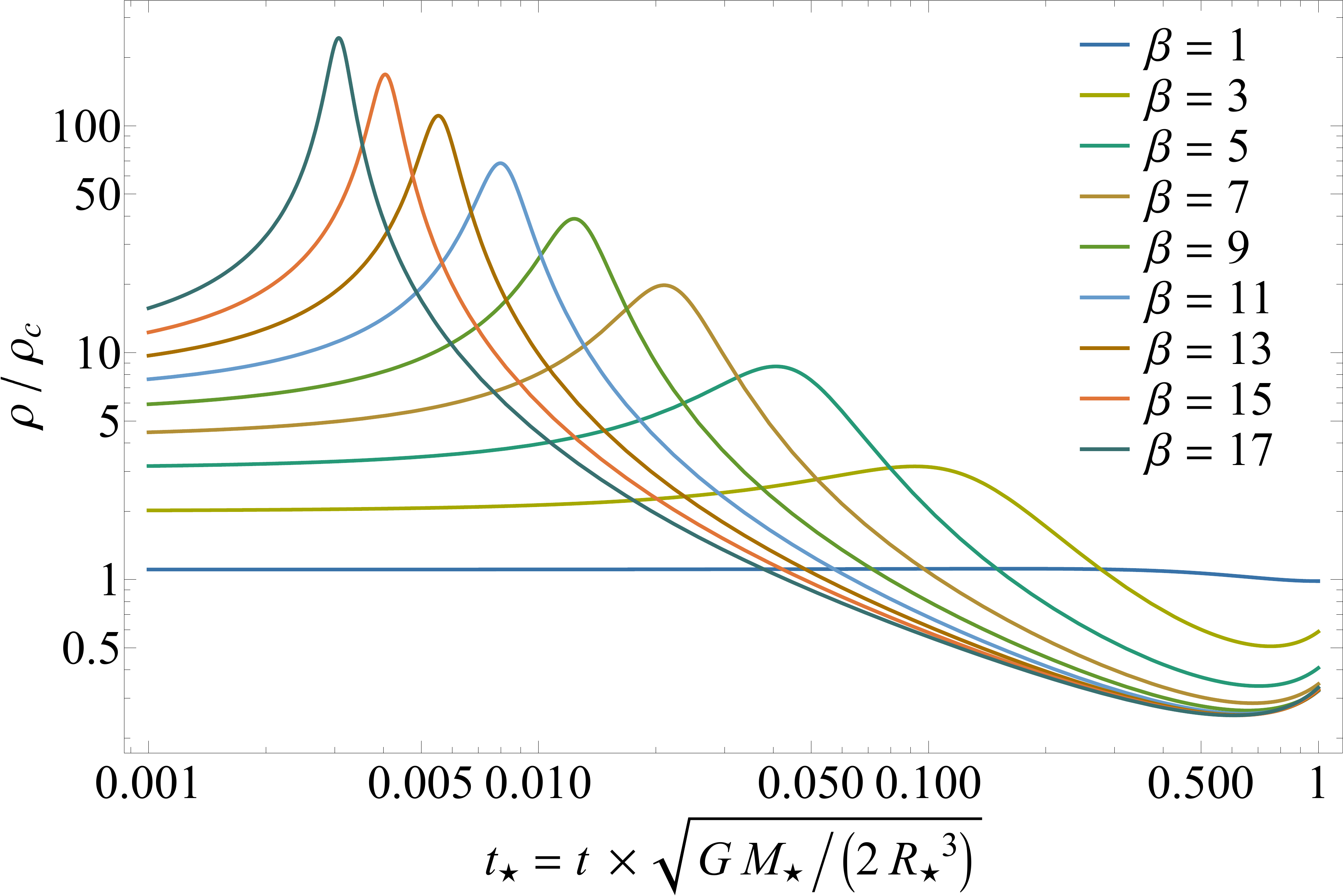} 
   \includegraphics[width=0.495\textwidth]{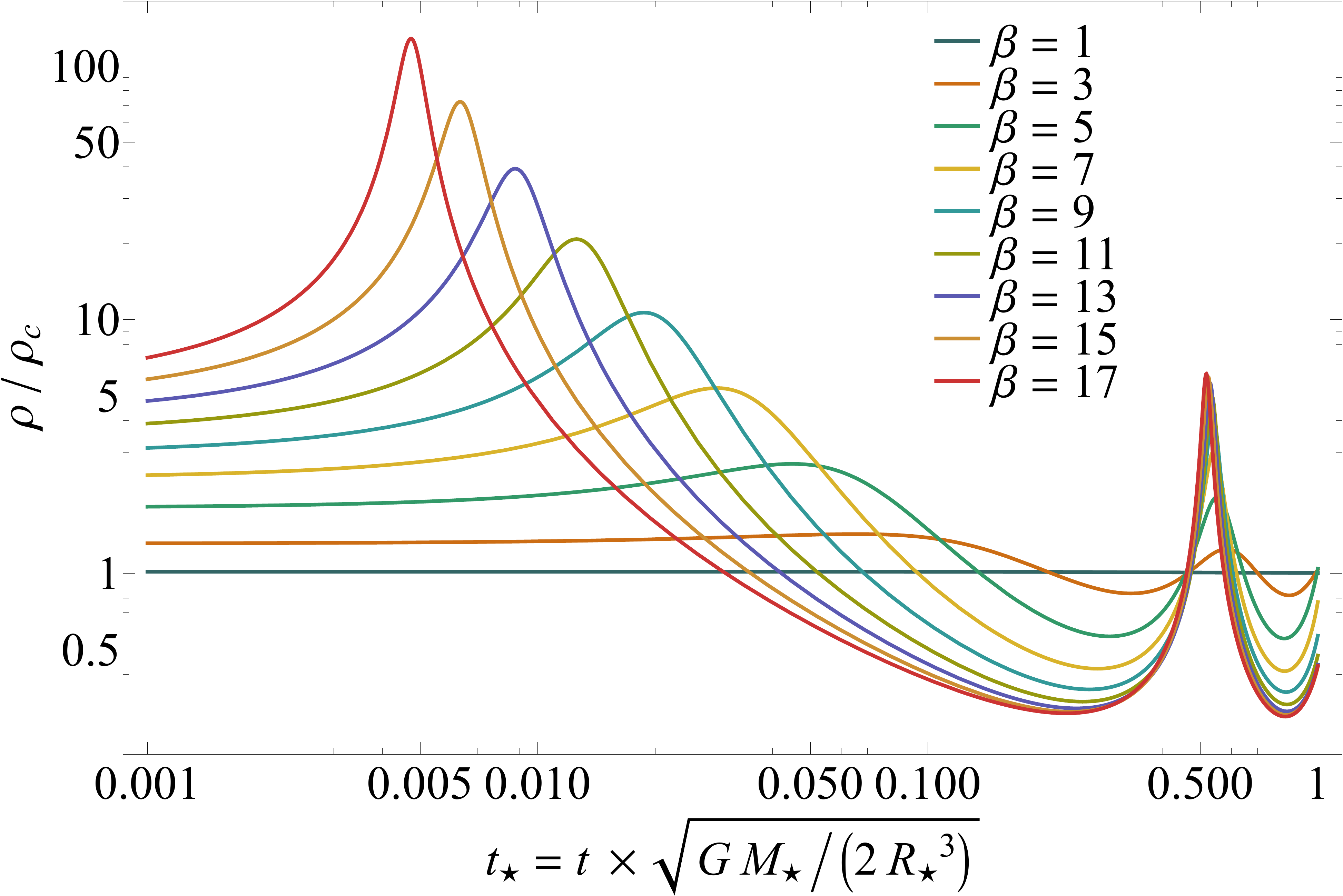} 
   \caption{The ratio of the central density to the original central density as a function of time normalized by the dynamical time of the star for the solution that incorporates pressure and self-gravity (i.e., the solution to Equation \ref{Hdyn}). The left (right) panel is for a $\gamma = 5/3$ ($\gamma = 4/3$) polytrope. We see that as $\beta$ increases the time at which the density is maximized approaches zero (which coincides with when the center of mass is at pericenter) and the maximum in the density increases. }
   \label{fig:H_of_t_lambda10_g53}
\end{figure*}

For a $\gamma = 5/3$ polytrope, $\rho_{\rm c}/\rho_{\star} \simeq 5.99$, while for a $\gamma = 4/3$ polytrope $\rho_{\rm c}/\rho_{\star} \simeq 54.2$. Because the term $\propto \rho_{\rm c}/\rho_{\star}$ is responsible for balancing the dynamical terms once $H$ is sufficiently small, this shows that more centrally concentrated stars are better able to withstand the tidal compression of the black hole. We also see that, for the same ratio of central to average stellar density, stars with smaller $\gamma$ will reach a smaller $H$ before gas pressure is able to reverse the infall because $H^{-\gamma}$ rises more slowly.

Figure \ref{fig:H_Hff} compares the solution for $H$ when pressure and self-gravity are ignored (dashed lines, given by Equation \ref{Heq}) and when they are included (solid lines, the solution to Equation \ref{Hdyn}). The left (right) panel is for a $\gamma = 5/3$ ($\gamma = 4/3$) polytrope. Time is in units of $\tau$; the pressureless solutions pass through the origin at a time of $\sinh(\tau_{\rm cr}) = \sqrt{\beta}-\sqrt{\beta-1}$, derivable from Equation \eqref{Heq}. We see, however, that pressure and self-gravity are important for modifying $H$ at much earlier times.  

\begin{figure*}[htbp] 
   \centering
   \includegraphics[width=0.495\textwidth]{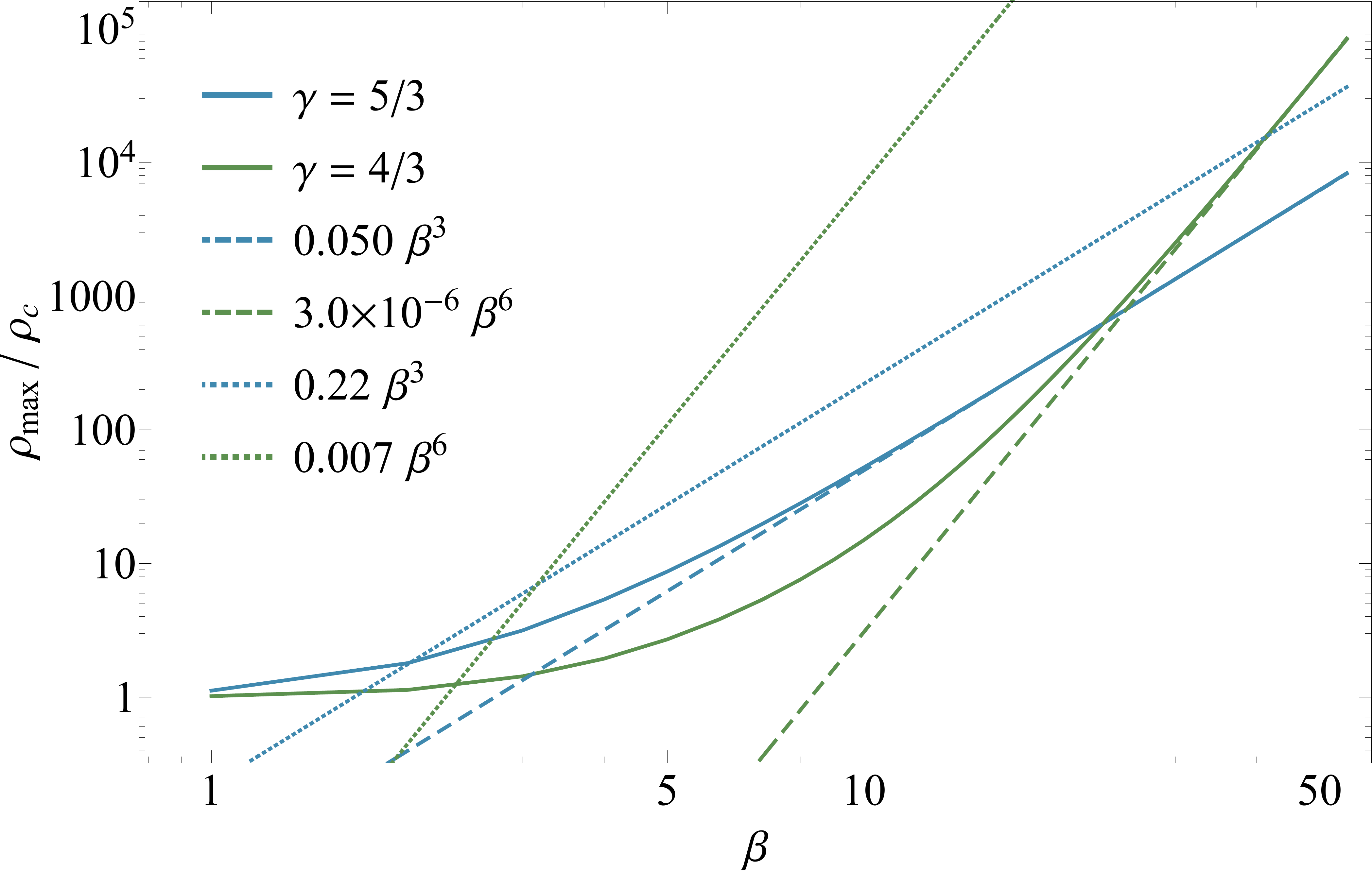} 
   \includegraphics[width=0.495\textwidth]{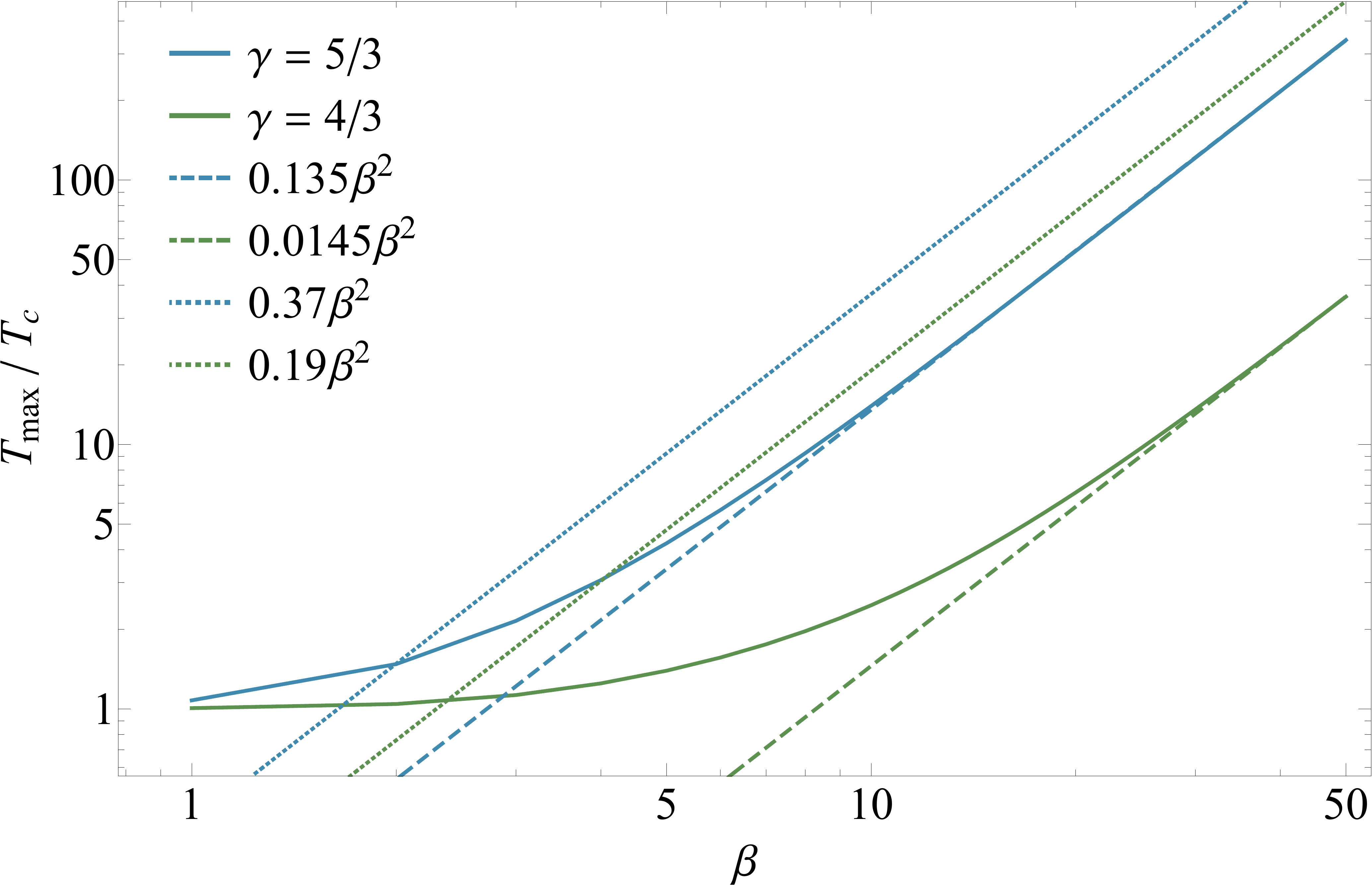} 
   \caption{The maximum density (left) and temperature (right) normalized by the central density and temperature of the star as a function of $\beta$ for $\gamma = 5/3$ (blue, solid) and $\gamma = 4/3$ (green, solid) for the solution that incorporates pressure and self-gravity (i.e., the solution to Equation \ref{Hdyn}). Fitted lines to the large-$\beta$ values are shown by the dashed lines, while the scalings predicted by \citet{carter83} for $\gamma = 5/3$ (dotted, blue) and $\gamma = 4/3$ (dotted, green) are also shown.}
   \label{fig:rhomax}
\end{figure*}

Figure \ref{fig:H_of_t_lambda10_g53} shows the density normalized by the central density of the star, which is just $H^{-1}$, as a function of time normalized by the dynamical time of the star for $\gamma = 5/3$ (left) and $\gamma = 4/3$ (right) and for the $\beta$ in the legend. As the $\beta$ of the encounter increases, the maximum density achieved by the star increases and occurs at an earlier time (note that the center of mass of the star reaches pericenter at $t = 0$). 

Figure \ref{fig:rhomax} illustrates the maximum central density (left) and the maximum central temperature (right) achieved during the tidal encounter as a function of $\beta$ for $\gamma = 5/3$ (solid, blue) and $\gamma = 4/3$ (solid, green); the temperature is calculated by assuming that gas-pressure dominates, such that $T \propto p/\rho$. The dashed lines show fits to the large-$\beta$ values of the maximum density, being $\propto \beta^3$ for the maximum density and $\gamma = 5/3$ (blue, dashed) and $\propto \beta^6$ for the density and for $\gamma = 4/3$ (green, dashed), while the maximum temperature scales as $\propto \beta^2$ for both $\gamma = 5/3$ and $4/3$. These scalings were also predicted by \citet{carter83} and \citet{luminet86}, which are shown by the dotted-blue and dotted-green lines; however, the coefficient of proportionality multiplying the scaling of the density is roughly a factor of five smaller than that predicted by \citet{luminet86} for $\gamma = 5/3$ and three orders of magnitude smaller for $\gamma = 4/3$, and a factor of $\sim 3$ smaller for the central temperature and $\gamma = 5/3$ and a factor of $\sim 10$ smaller for $\gamma = 4/3$. Interestingly, the scaling of the temperature and the normalization we recover from the homologous model is very similar to that derived by \citet{Bicknell:1983aa} (their Equation 2.14). However, this agreement is purely by coincidence: those authors argued that shock heating would alter the relationship predicted by \citet{carter83}, whereas here our reduced coefficient simply arises from the fact that the pressure gradient counteracts the tidal compression when the gas pressure is only a fraction of the freefall ram pressure. These large-$\beta$ scalings for the central density and temperature at the point of maximum compression can also be recovered from Equation \eqref{Hdyn} by assuming that $H$ follows its freefall solution -- and is very nearly zero -- until reaching the pericenter distance, at which point the second derivative is comparable to the pressure term so that the infall is reversed. 

\begin{figure*}[htbp] 
   \centering
   \includegraphics[width=0.495\textwidth]{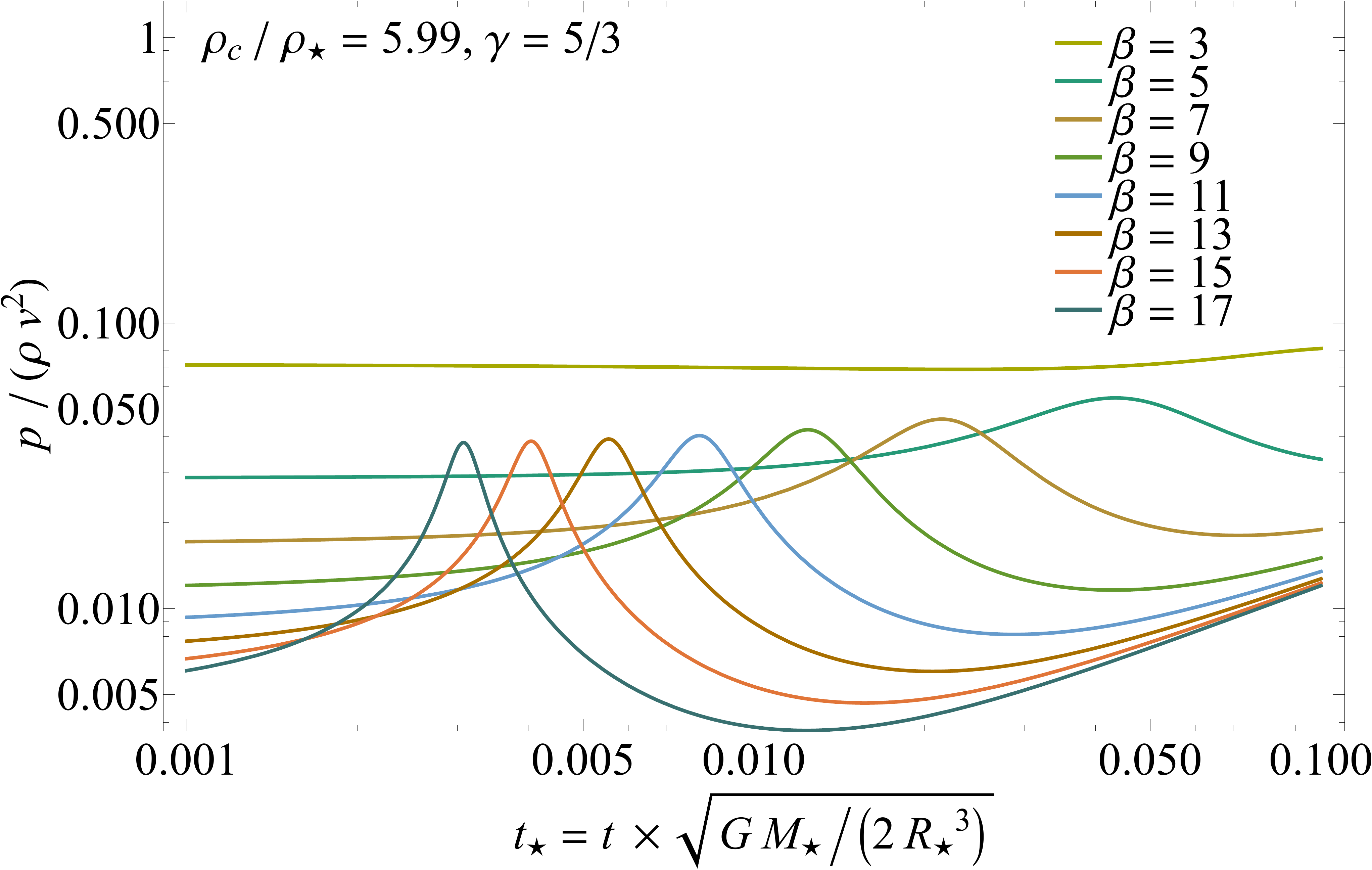} 
   \includegraphics[width=0.495\textwidth]{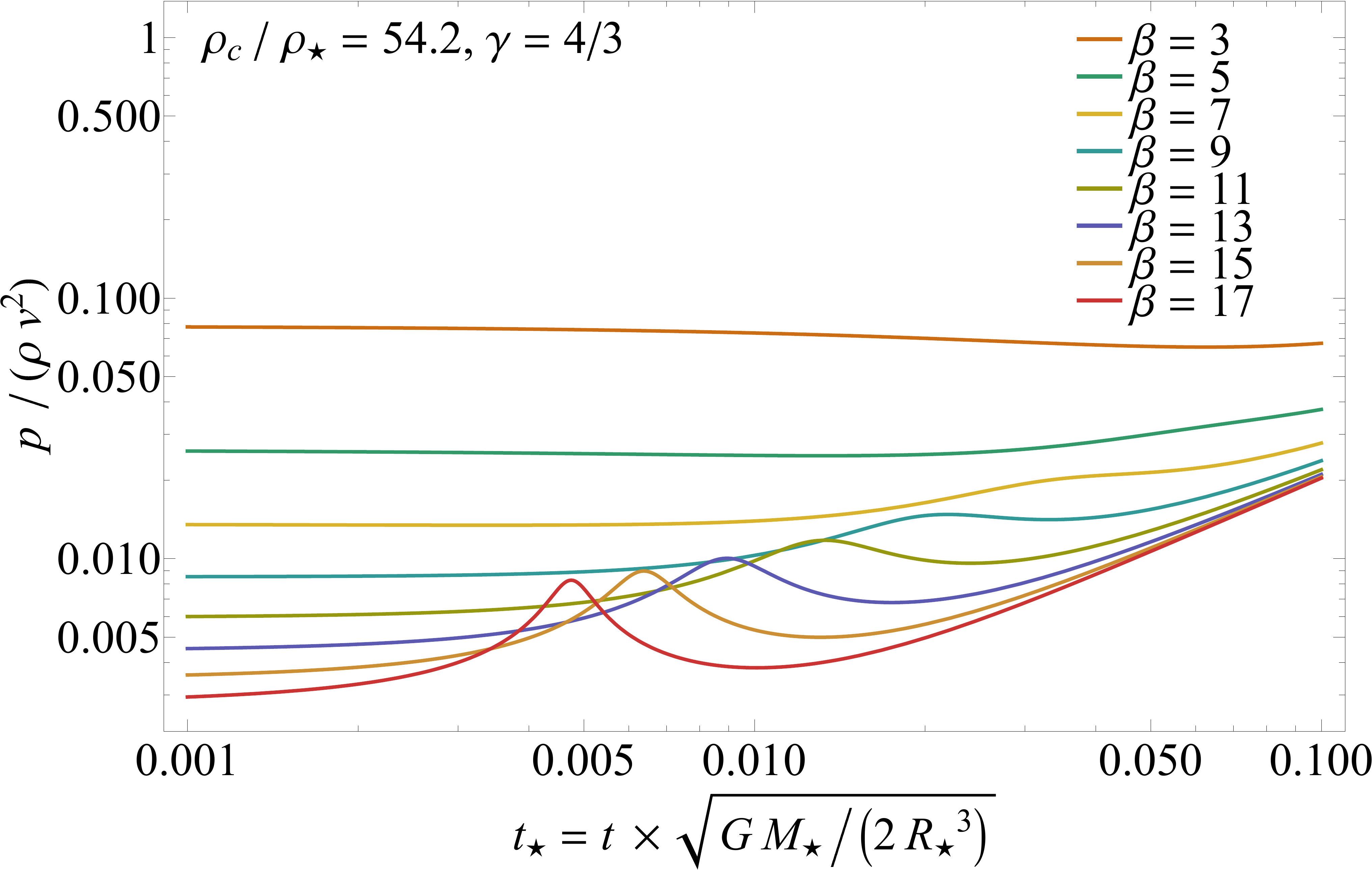} 
   \caption{The ratio of the gas pressure to the ram pressure, where the velocity is calculated from the freefall solution, for a $\gamma = 5/3$ polytrope (left) and a $\gamma = 4/3$ polytrope (right) for the solution that incorporates pressure and self-gravity (i.e., the solution to Equation \ref{Hdyn}). The fact that this ratio is much less than unity demonstrates that the pressure gradient is able to withstand the tidal compression well before the gas pressure is comparable to the ram pressure.}
   \label{fig:pgas_pram}
\end{figure*}

The fact that the coefficient multiplying the maximum density as a function of $\beta$ is so much smaller than the value predicted by \citet{carter83} and \citet{luminet86} (and similarly for the temperature) suggests that, in line with our reasoning in the previous section, the pressure gradient is withstanding the vertical compression when the gas pressure is much smaller than the ram pressure of the pressureless, freefalling material at the surface of the star. Figure \ref{fig:pgas_pram} shows the ratio of the gas pressure to the ram pressure of the pressureless solution for a $\gamma = 5/3$ polytrope (left) and a $\gamma = 4/3$ polytrope (right). For large $\beta$ this ratio approaches a constant near the time of peak compression (when each curve reaches a maximum), indicating that $\rho \propto \beta^{2/(\gamma-1)}$ is being upheld. However, the constant is much smaller than $\sim 1$.

\begin{figure}[htbp] 
   \centering
   \includegraphics[width=\columnwidth]{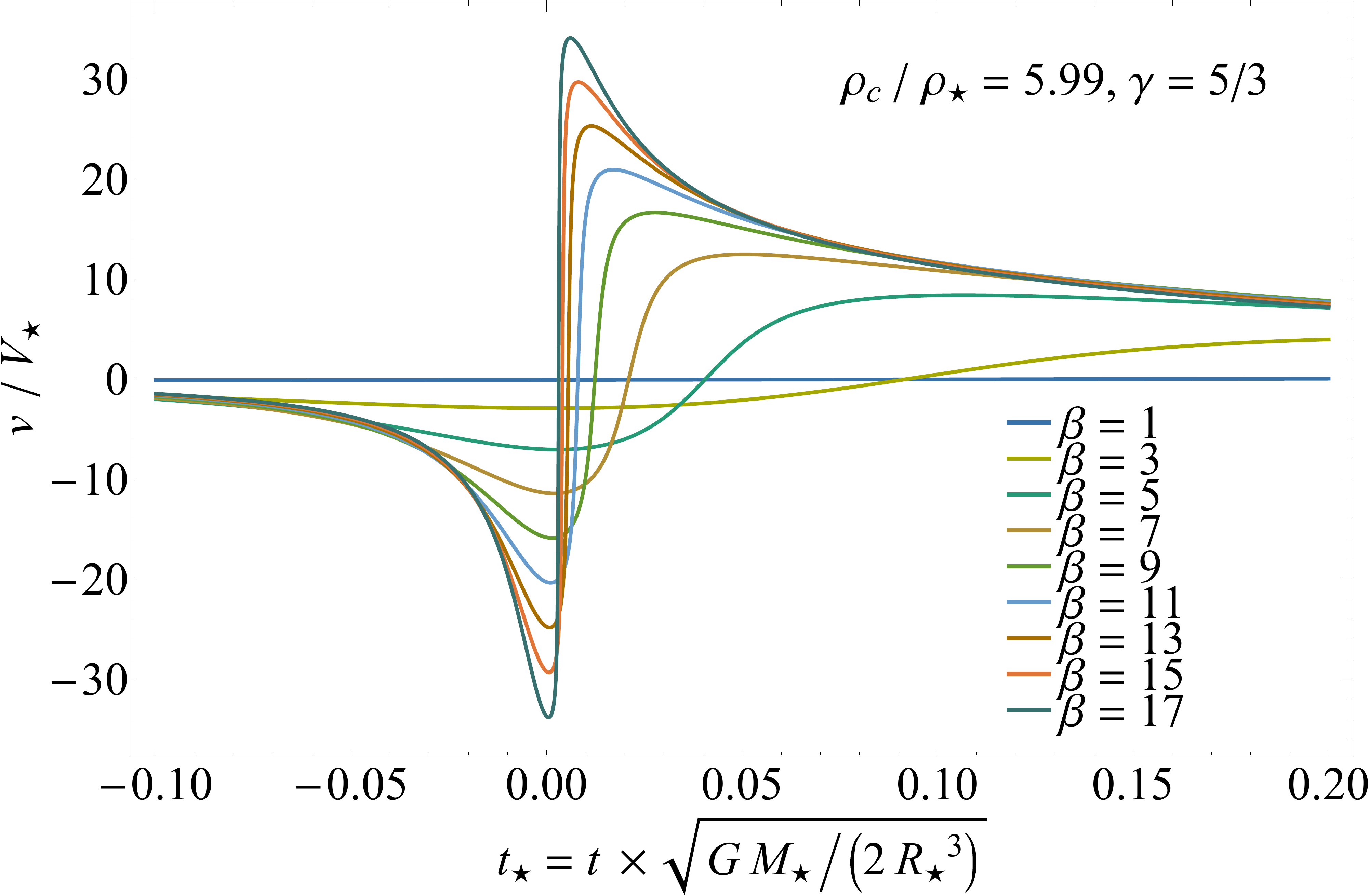}
   \caption{The velocity of the fluid element at the surface of the collapsing star normalized by the escape speed of the stellar progenitor as a function of time; here the progenitor is a $\gamma = 5/3$ polytrope and the solution incorporates pressure and self-gravity (i.e., the solution to Equation \ref{Hdyn}). The different curves are for the $\beta$ shown in the legend.}
   \label{fig:v_vesc}
\end{figure}

The velocity at the surface of the star is

\begin{equation}
v_{\rm z} = \frac{\partial z}{\partial t}(z_0 = R_{\star}) = V_{\star}\frac{\beta^{3/2}}{\cosh^3(\tau)}\dot{H}(\tau),
\end{equation}
where $V_{\star} = (\alpha/R_{\star})\sqrt{GM_{\star}/(2R_{\star})}$. Figure \ref{fig:v_vesc} shows this velocity as a function of time for a $\gamma = 5/3$ polytrope, where the different curves correspond to the $\beta$'s given in the legend. We see that the velocity becomes increasingly negative approximately until the star reaches pericenter, at which point it is minimized at $v \simeq -2\beta V_{\star}$ (which agrees with the freefall solutions; \citealt{Stone:2013aa}). The pressure then reverses the infall, causing the velocity to equal zero and rebound to reach a maximum velocity that is $v \simeq 2\beta V_{\star}$ before decaying again. 

Before moving on to the next section and the inclusion of non-homologous terms on the evolution of the compressing fluid in a tidal disruption event, we note that \citet{brassart08} concluded that the nearly homologous velocity profiles of their one-dimensional, hydrodynamical simulations of deep TDEs implied that the compression they observed was dynamical and that the effects of pressure and self-gravity were negligible. Our results here demonstrate that this conclusion is incorrect: compression can proceed homologously with the effects of pressure and self-gravity included, and in fact must be homologous to leading order in the relation between $z$ and $z_0$ -- whether or not pressure and self-gravity are included. The existence of a homologous velocity profile therefore provides no information as to the relevance, contribution, or importance of pressure or self-gravity to the dynamics of a fluid. Indeed, the kinetic energy is negligibly small -- and the dynamics still homologous -- near the point of maximum compression of the fluid. 

\newpage
\section{Non-homologous evolution}
\label{sec:non-homologous}
Our analysis in the preceding section\footnote{From this point on we restrict our analyses and discussion to gas-pressure dominated stars with $\gamma = 5/3$. We do so simultaneously for brevity and because radiation-pressure dominated stars are very massive and are short-lived and few in number, implying that the likelihood of disrupting one by a supermassive black hole (and observing it) is significantly reduced (though their extended envelopes make it somewhat more likely).} shows that the time-dependent positions of fluid elements are related to their initial positions by $z = H(\tau)z_0$, the same relationship that results from assuming that the gas freefalls in the tidal potential, but now $H$ is given by the solution to Equation \eqref{Hdyn}, which includes the effects of pressure and self-gravity. While this homologous relationship is exact (within the tidal approximation) for pressureless freefall, {it is only approximate} when we account for the effects of pressure and self-gravity. 
In particular, when deriving Equation \eqref{Hdyn} we equated linear terms in $z_0$, but we dropped higher-order terms. To satisfy the momentum equation to the next order in $z_0$, we must therefore have

\begin{equation}
z = H_{00}(\tau)z_0+H_{10}(\tau) z_0^3 + H_{01}(\tau)z_0s_0^2. \label{znonhomo}
\end{equation}

We can now follow the same steps as we did in the previous subsection to derive three equations for $H_{00}$, $H_{10}$, and $H_{01}$ by equating first-order terms in $z_0$, terms proportional to $z_0s_0^2$, and and third-order terms in $z_0$ in the momentum equation. The expression for the density that follows from mass conservation is still upheld, namely

\begin{equation}
\rho = \rho_{\rm c}\frac{\partial z_0}{\partial z}g(s_0,z_0),
\end{equation}
and we now include the next-higher-order term in the series expansion for the density about the origin:

\begin{equation}
g_0(s_0,z_0) = 1-\left(s_0^2+z_0^2\right)+\frac{13-5\gamma}{10}\left(s_0^2+z_0^2\right)^2. \label{rho0}
\end{equation}
The last term was derived from the equation of hydrostatic balance. From Equation \eqref{znonhomo}

\begin{equation}
\frac{\partial z_0}{\partial z} = \frac{1}{H_{00}+H_{01}s_0^2+3H_{10} z_0^2}, 
\end{equation}
and the pressure is still given by $p = K\rho^{\gamma}$. The gravitational potential\footnote{We ignore the octupole term in the gravitational potential of the black hole because it is weaker than the other terms by a factor of $\sim (M_{\bullet}/M_{\star})^{-2/3}\beta^2$, and is therefore only important when $\beta \gtrsim 100$. For $\beta$'s this large, relativistic effects, nuclear fusion, etc., are much more important and invalidate our Newtonian and purely hydrodynamical treatment here.} satisfies the Poisson equation, which to this order is

\begin{multline}
\nabla^2\Phi = 4\pi G\rho_{\rm c}\frac{\partial z_0}{\partial z}\left(1-s_0^2-z_0^2\right) \\
\simeq \frac{4\pi G\rho_{\rm c}}{H_{00}}\left(1-\left(1+\frac{H_{01}}{H_{00}}\right)s_0^2-\left(1+\frac{3H_{10}}{H_{00}}\right)z_0^2\right),
\end{multline}
Transforming to spherical coordinates and writing $z_0$ in terms of $z$, the solution to this equation is

\begin{equation}
\Phi = 4\pi G\rho_{\rm c}\alpha^2H_{00}^{-1}\frac{1}{6}r^2\left\{1-\frac{3A}{10}r^2+\frac{3B}{7}r^2Y^{0}_2(\theta)\right\}, \label{Phisol}
\end{equation}
where

\begin{equation}
A = \frac{1}{3}\left(2+2\frac{H_{01}}{H_{00}}+\frac{1}{H_{00}^2}+3\frac{H_{10}}{H_{00}^3}\right),
\end{equation}
\begin{equation}
B = \frac{1}{3}\left(1+\frac{H_{01}}{H_{00}}-\frac{1}{H_{00}^2}-\frac{3H_{10}}{H_{00}^3}\right),
\end{equation}
and

\begin{equation}
Y^{0}_{2}(\theta) = 3\cos^2\theta-1
\end{equation}
is the $\ell = 2$, $m = 0$ spherical harmonic. 

We can now insert our solutions for the gravitational potential, the pressure, the density, and $z(z_0,\tau)$ into the $z$-momentum equation, Taylor expand, and equate terms in $z_0$, $z_0s_0^2$, and $z_0^3$. The result is a set of three equations for $H_{00}$, $H_{10}$, and $H_{01}$ that are of the form

\begin{equation}
\mathcal{L}\left[H_{\rm i}\right]+\frac{2\rho_{\rm c}}{\rho_{\star}}\frac{\cosh^6(\tau)}{\beta^3}F_{\rm i}\left[H_{\rm i}\right]= 0, \label{H00nl}
\end{equation}
where $H_{\rm i}$ is $H_{00}$, $H_{10}$, or $H_{01}$ and $F_{\rm i}$ is a function of the $H$'s. For example, the equation for $H_{00}$ is

\begin{equation}
\mathcal{L}\left[H_{00}\right]+\frac{2\rho_{\rm c}}{\rho_{\star}}\frac{\cosh^6(\tau)}{\beta^3}\left\{1-H_{00}^{-\gamma}\left(1+\frac{3H_{10}}{H_{00}}\right)\right\} = 0.
\end{equation}
The initial conditions on $H_{00}$ are, as before, $H_{00}(\tau \rightarrow -\infty) = 1$ and $\dot{H}_{00}(\tau \rightarrow -\infty) = 0$, while both $H_{10}$ and $H_{01}$ and their derivatives are zero at $\tau \rightarrow -\infty$. An important point to note is that all of the dynamical equations are {coupled}: the equation for $H_{00}$ is no longer exactly what we derived in Section \ref{sec:homologous} (Equation \ref{Hdyn}), which ignored the nonlinear contribution to the dynamics, but now -- as can be seen from Equation \eqref{H00nl} -- includes a term that depends on $H_{10}$. This feature demonstrates that the central density of the star is affected by the nonlinearity of the solution.

\begin{figure*}[htbp] 
   \centering
   \includegraphics[width=0.495\textwidth]{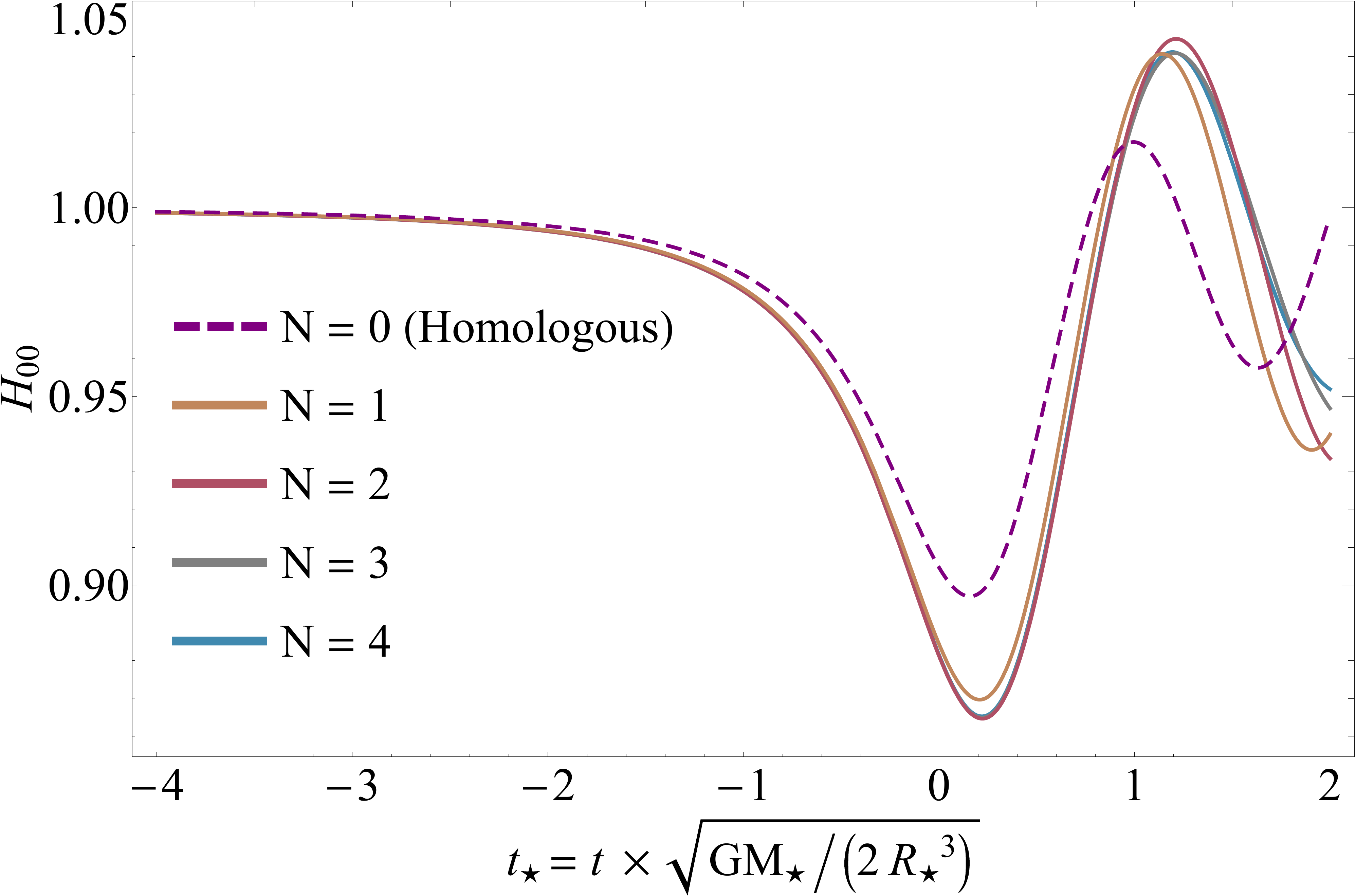} 
   \includegraphics[width=0.495\textwidth]{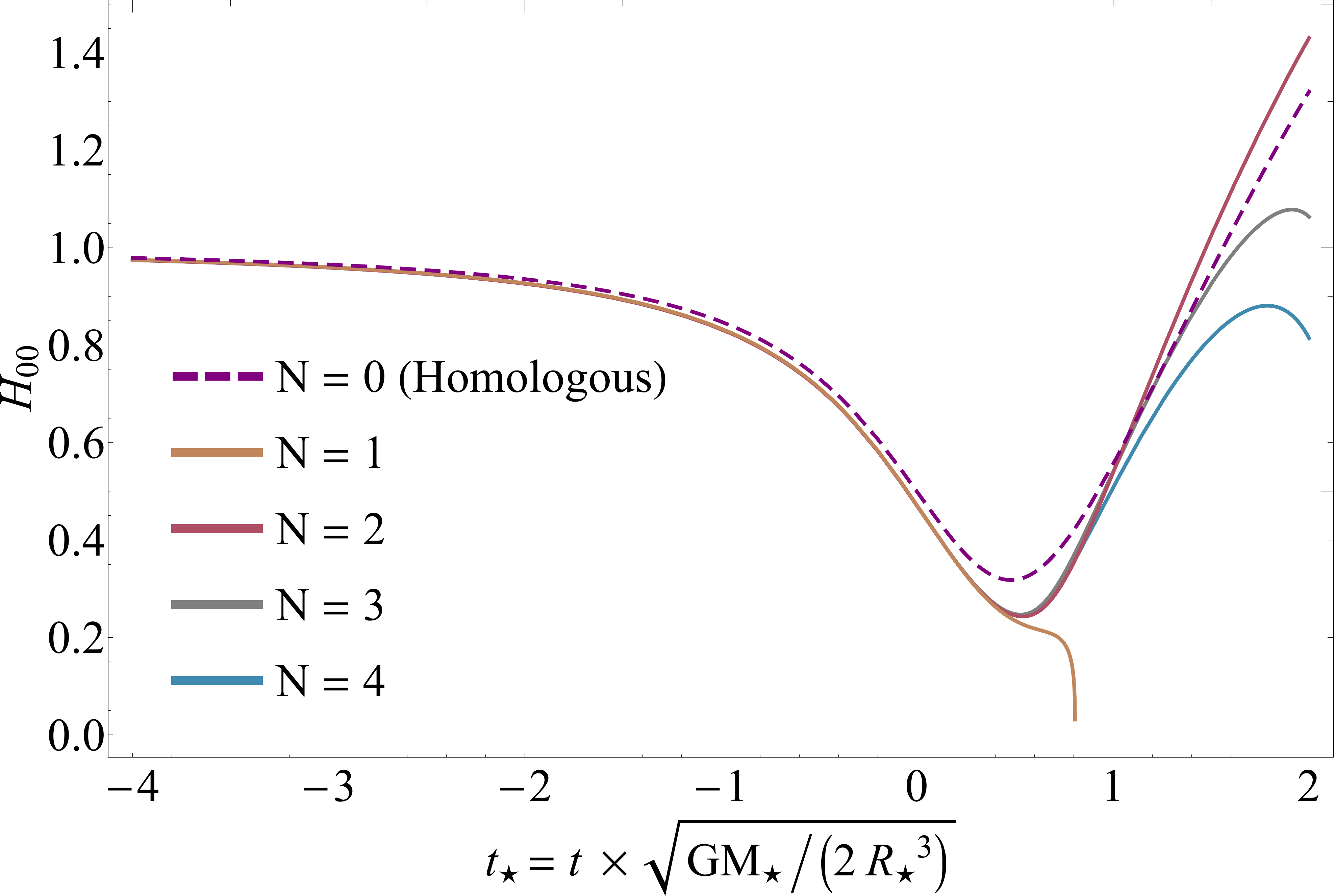} 
   \caption{The function $H_{00}$, which is roughly the ratio of the current to the initial Lagrangian height of a fluid element, and the inverse of which is the density at the center of the star relative to its initial density, for $\beta = 1$ (left) and $\beta = 3$ (right) for a $\gamma = 5/3$ polytrope. The different curves are appropriate to different levels of nonlinearity in the relationship between $z$ and $z_0$, with the highest term in the expansion being $z_0^{2N+1}$. We see that the higher-order solutions lie on top of one another out to a time that is $\sim$ 1 dynamical time of the star (for times much greater than this our solution breaks down as it does not capture the in-plane spreading of the debris, which results in a decrease in the density, or an increase in $H_{00}$, and the complete disruption of the star).}
   \label{fig:H00_comps}
\end{figure*}

Extending the solutions to higher order is straightforward: we decompose $z$ in terms of $s_0$ and $z_0$ as

\begin{equation}
z = \sum_{j = 0}^{N} \sum_{n+m= j} H_{\rm nm}(\tau)s_0^{2m}z_0^{2n+1}. \label{zofz0}
\end{equation}
The bound of $n+m = j$ means over all combinations of $n$ and $m$ such that $n+m = j$. For example, including the fifth-order term in $z_0$ gives

\begin{equation*}
z = H_{00}z_0+H_{10}z_0^3+H_{01}s_0^2z_0+H_{20}z_0^5+H_{11}s_0^2z_0^3+H_{02}s_0^4z_0.
\end{equation*}
For a maximum exponent of $z_0$ of $2N+1$, there are a total of $(N+1)(N+2)/2$ equations for the $H_{\rm nm}$. To determine the self-gravitational potential we expand the density to order $2N$ in $z_0$ and $s_0$, then substitute $z_0(z,s_0)$ (i.e., the inverse of Equation \ref{zofz0} to leading self-consistent order) to write the density in terms of the current coordinates $z$ and $s_0$. We then transform the density to spherical coordinates and write the gravitational potential as a sum of spherical harmonics, transform back to $z$ and $s_0$ from $r$ and $\theta$, differentiate with respect to $z$, and replace $z\rightarrow z(z_0,s_0)$ with Equation \eqref{zofz0}. We similarly expand $1/\rho\times\partial p/\partial z$ in terms of $z_0$ and $s_0$, insert the result, the derivative of the gravitational potential, and Equation \eqref{zofz0} into the $z$-momentum equation and equate equal powers of $s_0^{2m}z_0^{2n+1}$ to generate $(N+1)(N+2)/2$ equations for the $H_{\rm nm}(\tau)$. The hydrostatic initial conditions are that $H_{00}(\tau \rightarrow -\infty) = 1$, $\dot{H}_{00}(\tau \rightarrow -\infty) = 0$, and all of the other functions and their derivatives equal zero at $\tau \rightarrow -\infty$. 

Figure \ref{fig:H00_comps} shows $H_{00}$, which is the inverse of the density at the center of the star and is approximately the ratio of the current Lagangian height of a fluid element to its initial height, for $\beta = 1$ (left) and $\beta = 3$ (right). The different curves give the maximum number of terms in the expansion between $z$ and $z_0$, with the largest term in the expansion being $\propto z_0^{2N+1}$. We see that the homologous solution slightly underestimates the amount of compression that occurs during the tidal encounter. For $\beta = 1$, all of the curves lie effectively on top of one another until $\sim 1$ dynamical time of the star; for times beyond this our solutions do not accurately capture the evolution of the TDE because we have not included the in-plane motion of the fluid, which (after about the time at which the star recedes beyond the tidal radius on its egress; \citealt{coughlin20}) results in a continued decline of the density. 

\begin{figure*}[htbp] 
   \centering
   \includegraphics[width=0.495\textwidth]{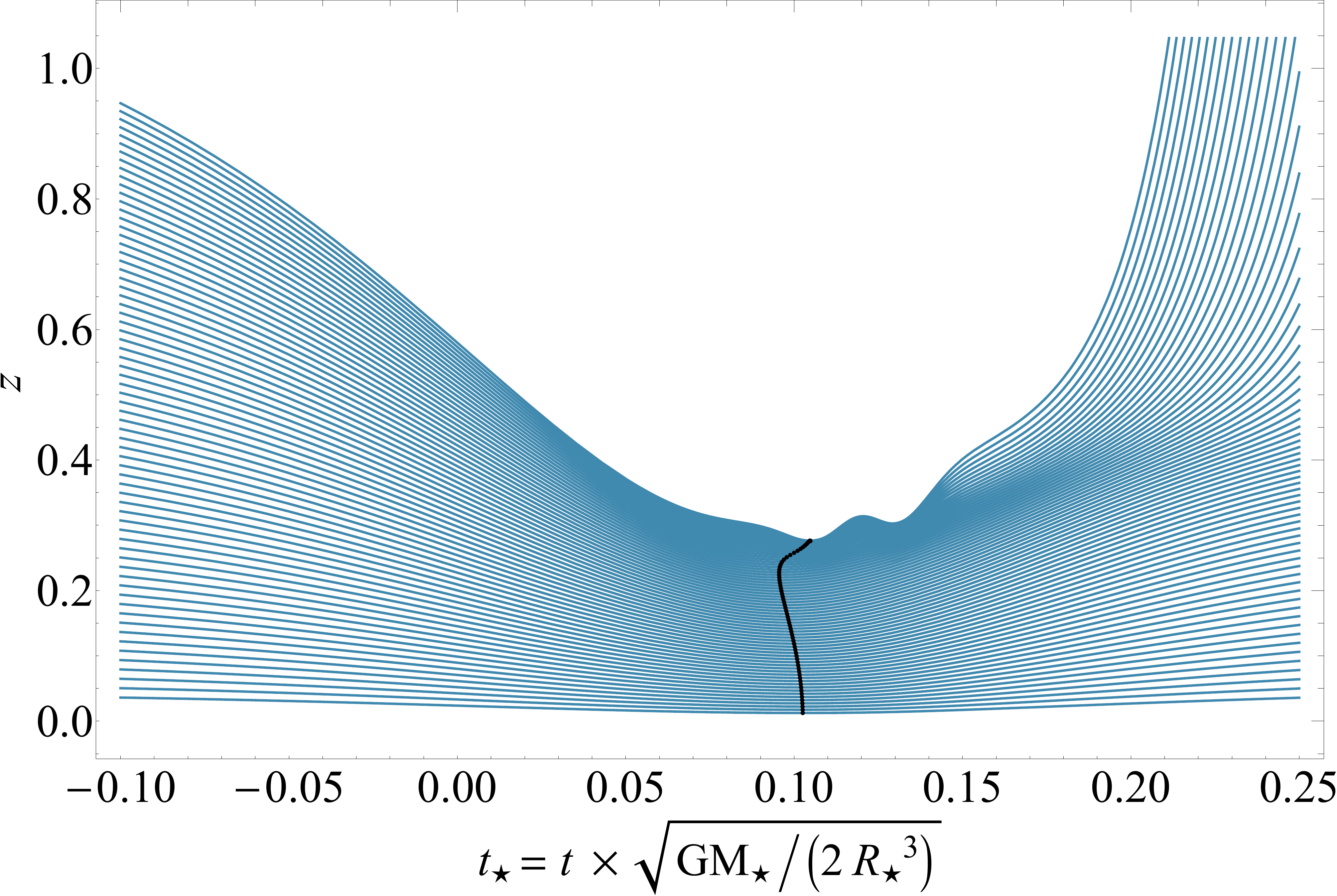} 
 \includegraphics[width=0.495\textwidth]{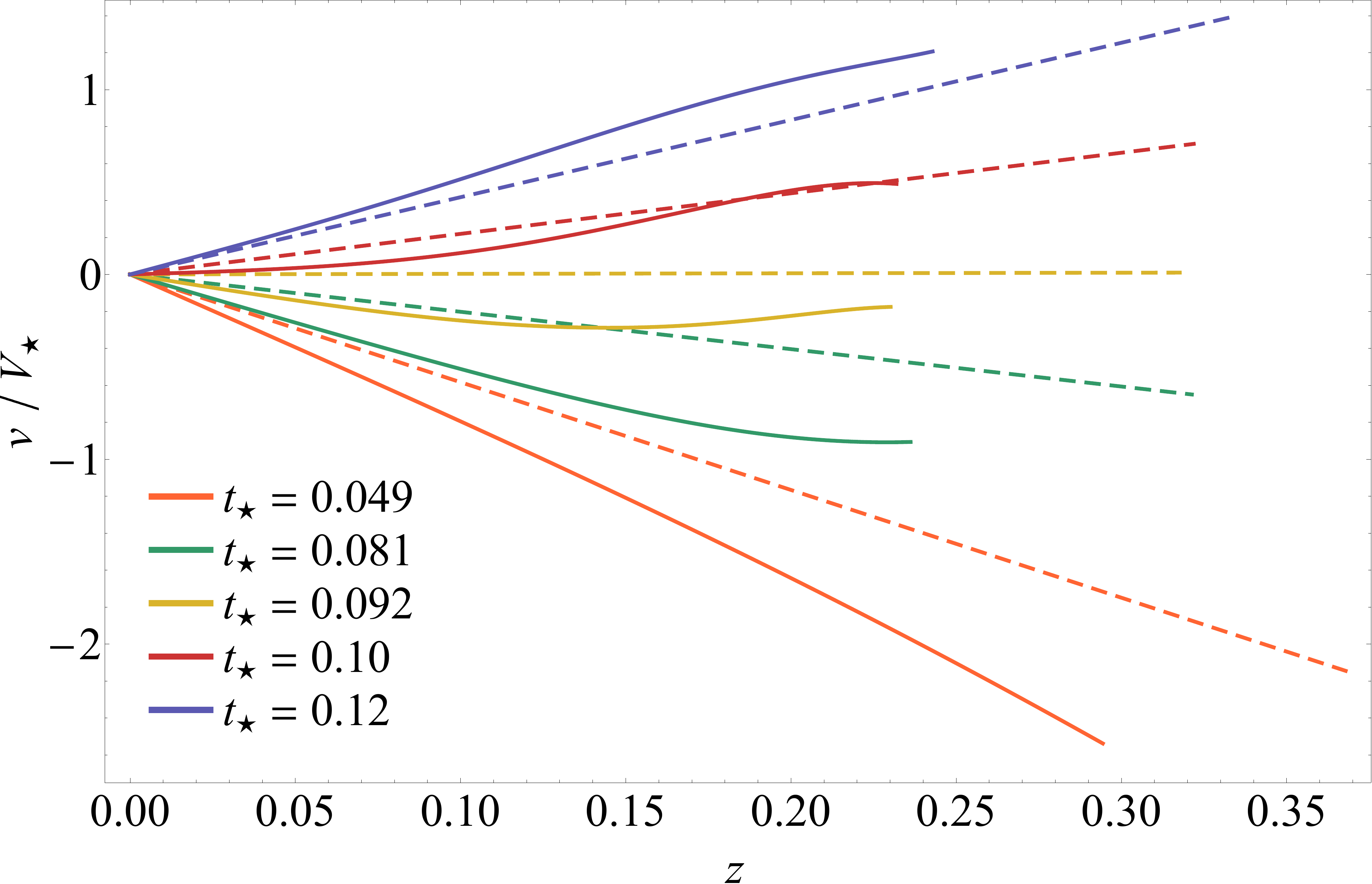} 
   \caption{Left: The Lagrangian height of each fluid element, $z$ (recall that $z$ is measured relative to $\alpha$, the scale height appropriate to the center of the star), at a cylindrical radius of $s_0 = 0$ (i.e., directly above the center of the star; blue curves) as a function of time in units of the dynamical time of the star for $\beta = 3$ when the disrupted star is a $\gamma = 5/3$ polytrope. The black points coincide with the time of maximum compression of each fluid parcel, and the fact that they do not lie on a vertical line shows that the collapse of the star possesses some degree of non-homology. Right: The Eulerian velocity profile normalized by $V_{\star} = \alpha\sqrt{GM_{\star}/(2R_{\star}^3)}$ (roughly the escape speed of the star) at the times shown in the legend for $\beta = 3$. The dashed curves are from the homologous solution, while the solid curves are from the nonlinear solution with $N = 4$. The nonlinear behavior of the velocity with height above the plane is a second illustration of the non-homologous nature of the stellar compression near pericenter. (Note that the pericenter of the center of mass is reached at $t_{\star} =0$.)}
   \label{fig:z_lag}
\end{figure*}

Figure \ref{fig:z_lag} shows two manifestations of the non-homologous nature of the evolution of the fluid when nonlinear terms are included in the expansion of $z(s_0,z_0)$: each curve in the left panel gives the Lagrangian height of a fluid element as a function of time in units of the dynamical time of the star for $\beta = 3$ and with $N = 4$. Here we let $s_0 = 0$ so that all the fluid parcels are directly above the center of the star. The black points near a time of $t_{\star} \simeq 0.1$ give the time of maximum compression of each fluid element, i.e., when each curve reaches a relative minimum. The fact that these points do not lie on a vertical line illustrates the non-homologous nature of the compression -- that different fluid elements reach their relative minima at different times. The right panel of this figure gives the Eulerian velocity profile of the fluid, where each curve is at the time shown in the legend. The dashed curves correspond to the homologous solution and satisfy $v \propto z$. The solid curves are from the nonlinear solution with $N = 4$, and it is apparent that these profiles possess nonlinear variation with their height above the plane. 

\begin{figure}[htbp] 
   \centering
   \includegraphics[width=0.475\textwidth]{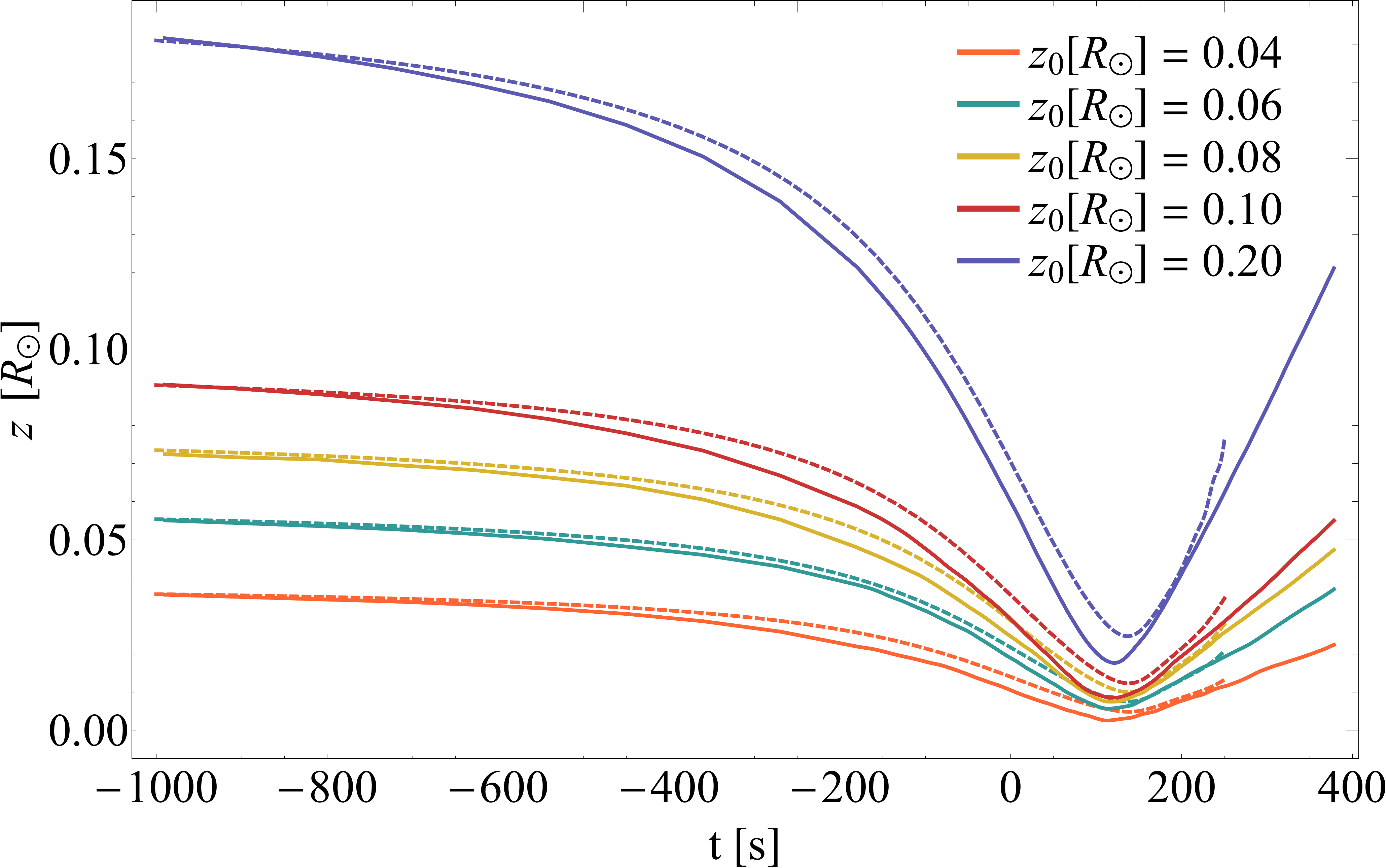} 
   \caption{The Lagrangian height of five fluid elements in units of solar radii as functions of time in seconds, where the initial height of each fluid element is shown in the legend, for a $\beta = 4$ disruption of a solar-like, $\gamma = 5/3$ polytrope. The dashed curves give the solution from the analytic model with $N = 4$, while the solid curves result from a numerical, SPH simulation performed with the SPH code {\sc phantom}. Here $\sim$ 128 million particles were used to model the disruption with {\sc phantom}.}
   \label{fig:lag_simulation_comps}
\end{figure}

Figure \ref{fig:lag_simulation_comps} illustrates the evolution of the Lagrangian height of five fluid elements directly above the center of the star ($s_0 = 0$), their initial heights shown in the legend, as functions of time in seconds for the disruption of a solar-like, $\gamma = 5/3$ polytrope (i.e., one with a solar mass and radius) when $\beta = 4$. The dashed curves are from the nonlinear solution with $N = 4$. The solid curves are from an SPH simulation performed with the {\sc phantom} \citep{Price:2018aa} SPH code. The setup of the simulation was identical to what is described in \citet{Coughlin:2015aa} (see \citealt{Miles:2020aa} for a more recent implementation), the most important aspects of the simulation being that the star was initialized in hydrostatic balance at a distance of $5r_{\rm t}$ and the equation of state was adiabatic. Here we used roughly 128 million particles; for more details of these specific simulations and more analysis thereof see \citet{norman21}. We see that there are some small differences between the predictions and the simulations; for example, the minimum height reached by each of the fluid elements from the simulation is slightly smaller than the model predicts, and the time at which the relative minimum is reached for each particle is slightly earlier than that from the analytic solution. Overall, however, the agreement is quite good.

So far we have ignored the in-plane motion and tidal stretching of the star. We can approximately account for this aspect of the disruption by assuming that the in-plane motion is ballistic, which leads to the following two equations of motion for the $x$ and $y$-positions of each fluid element relative to the center of mass (i.e., the $x$ position of a fluid element is $x+x_{\rm c}$, where $x_{\rm c}$ is the $x$-position of the center of mass):

\begin{equation*}
\ddot{x}-3\tanh(\tau)\dot{x}+2x = 6\cos\left(\varphi_{\rm c}\right)\left\{x\cos\left(\varphi_{\rm c}\right)+y\sin\left(\varphi_{\rm c}\right)\right\},
\end{equation*} 
\begin{equation*}
\ddot{y}-3\tanh(\tau)\dot{y}+2y = 6\sin\left(\varphi_{\rm c}\right)\left\{x\cos\left(\varphi_{\rm c}\right)+y\sin\left(\varphi_{\rm c}\right)\right\}.
\end{equation*} 
Here $\varphi_{\rm c}$ is the angle between the direction of pericenter of the stellar center of mass and the current position of the center of mass and is

\begin{equation}
\varphi_{\rm c} = 2\arctan\left[\sinh\left(\tau\right)\right].
\end{equation}
Assuming that the star retains hydrostatic balance prior to reaching the tidal radius gives the initial conditions $x(\tau_{\rm t}) = x_0$, $y(\tau_{\rm t}) = y_0$, $\dot{x}(\tau_{\rm t}) = \dot{y}(\tau_{\rm t}) = 0$, where (from Section \ref{sec:pressureless}) $\sinh(\tau_{\rm t}) = -\sqrt{\beta-1}$. The density at the center of the star is then approximately given by

\begin{equation}
\rho(x_0 = y_0 = z_0 = 0) = \rho_{\rm c}H_{00}(\tau)^{-1}D(\tau)^{-1}, \label{rhoinplane}
\end{equation}
where

\begin{equation}
D = \frac{\partial x}{\partial x_0}\frac{\partial y}{\partial y_0}-\frac{\partial x}{\partial y_0}\frac{\partial y}{\partial x_0}\bigg{|}_{x_0 = y_0 = 0}.
\end{equation}
Equation \eqref{rhoinplane} is only approximate in the sense that we have not accounted for the back reaction of the in-plane stretching on the evolution of $H_{00}(\tau)$. However, if the motion out of the plane is the predominant contributor to changes in the density, then Equation \eqref{rhoinplane} should accurately account for the relatively small changes in the density that are accompanied by the in-plane motion.

\begin{figure}[htbp] 
   \centering
   \includegraphics[width=0.475\textwidth]{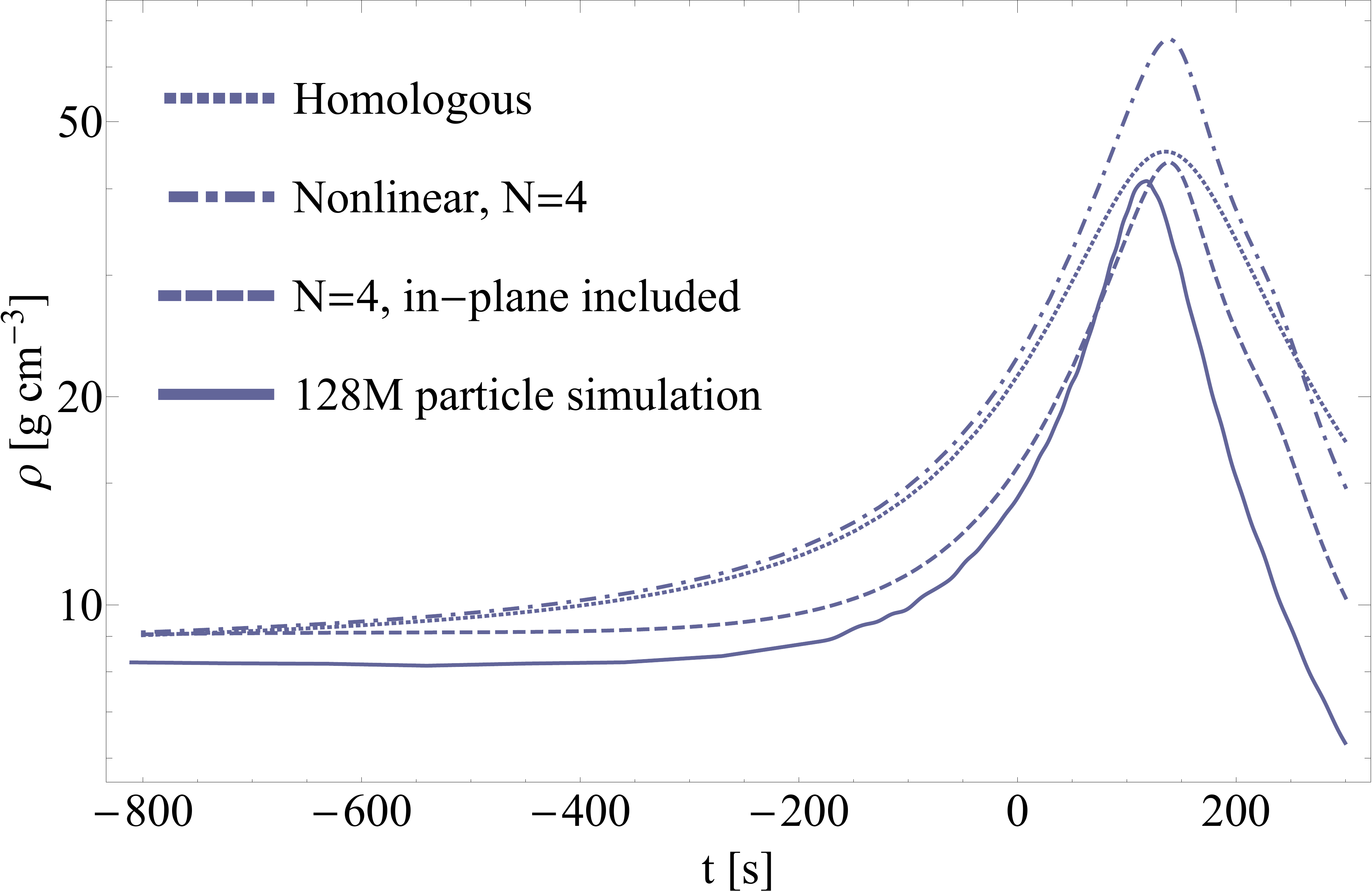} 
   \caption{The density as a function of time in seconds for the $\beta = 4$ disruption of a solar-like, $\gamma = 5/3$ polytrope. The dotted curve is the homologous solution, the dot-dashed curve is the $N=4$, nonlinear solution, the dashed curve is the $N = 4$, nonlinear solution with the in-plane expansion effects included, and the solid curve is the result of a 128 million-particle SPH simulation with {\sc phantom}. We see that the analytical model -- even at the homologous level -- provides a very good fit to the numerical results. }
   \label{fig:rho_comps}
\end{figure}

Figure \ref{fig:rho_comps} shows the central density (in cgs) as a function of time in seconds for a $\beta = 4$ disruption of a $\gamma = 5/3$, solar-like polytrope. The dotted curve results from the homologous approximation, the dot-dashed curve results from the $N = 4$, nonlinear solution, the dashed curve is the same solution but multiplied by the in-plane stretching factor, and the solid curve is the data from the 128 million-particle, SPH simulation run with {\sc phantom}. We see that accounting for the in-plane motion brings the analytic prediction in very close agreement with the hydrodynamical simulation, while ignoring this aspect of the evolution results in an overestimate. Coincidentally, the homologous solution very accurately reproduces the maximum density and the time of maximum; the reason for this agreement is that the nonlinear terms increase the density above the homologous value, while the in-plane stretching reduces the density by roughly the same value. 

\begin{figure}[htbp] 
   \centering
   \includegraphics[width=0.475\textwidth]{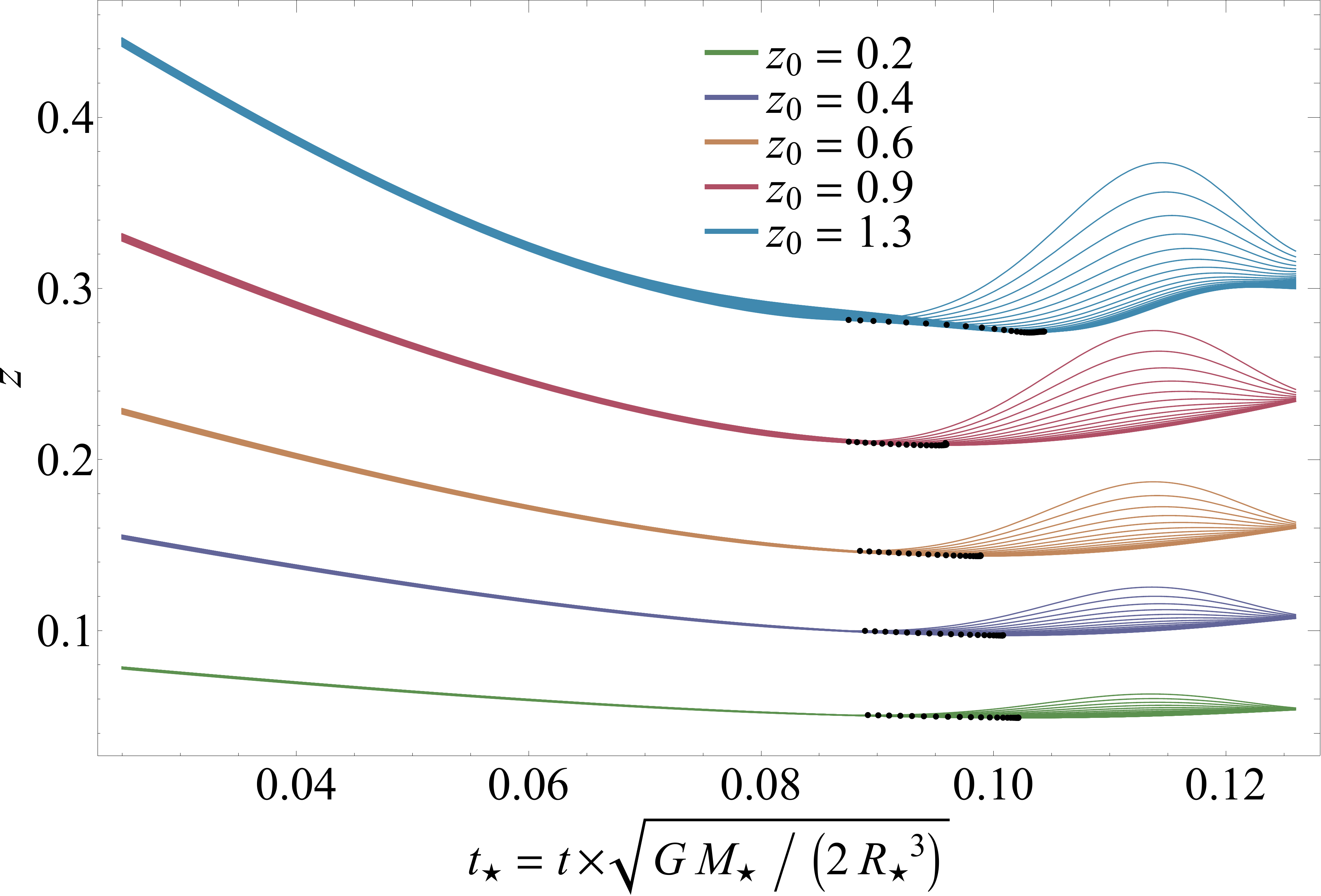} 
   \caption{The Lagrangian height above the plane for fluid elements at different initial heights, where the initial heights are given in the legend, for a $\gamma = 5/3$ polytrope and $\beta = 3$; here the solution was obtained with $N = 4$ (maximum term in the expansion between $z$ and $z_0$ $\propto z_0^{9}$). The different curves are for different initial cylindrical distances from the center of the star, where the bottom-most curve for each set of colored lines has $s_0 = 0$ (above the center of the star), while the top-most curve has $s_0 = 0.73$ (recall that $s_0$ and $z_0$ are measured relative to $\alpha$, the central scale height of the star). The black points illustrate the relative minimum of each fluid element.}
   \label{fig:z_lag_s0}
\end{figure}

In addition to depending on the initial height above the plane, the non-homologous relationship between $z$ and $z_0$ also depends on the initial cylindrical radius of the fluid element, $s_0$. Figure \ref{fig:z_lag_s0} shows the Lagrangian height above the plane $z$ as a function of time in units of the dynamical time of the star for the disruption of a $\gamma = 5/3$ polytrope with $\beta = 3$ and the solution with $N = 4$, so the maximum exponent in the relationship between $z$ and $z_0$ is $\propto z_0^{9}$. The different colors are appropriate to different initial heights above the plane, as shown in the legend, and the different curves within each colored set are for a different initial cylindrical radius; the bottom-most curve has $s_0$ = 0 (directly above the center of mass) while the top-most curve has $s_0 = 0.73$. The black points show the relative minimum of each curve. This figure illustrates that the compression of the star is a multidimensional process, and that fluid elements at different heights and cylindrical radii reach their relative minima at different times and decompress at different rates. 

\begin{figure}[htbp] 
   \centering
   \includegraphics[width=0.475\textwidth]{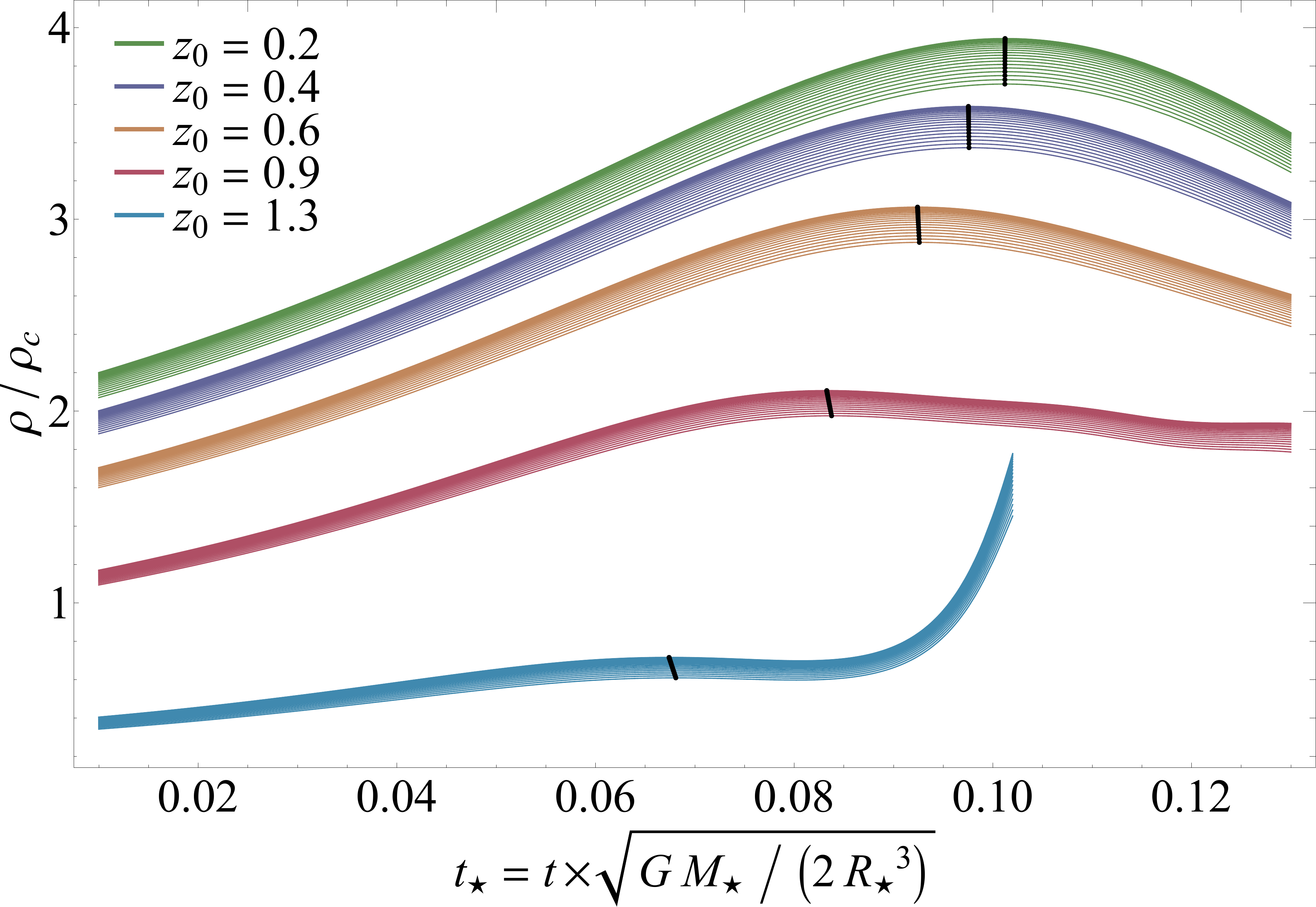} 
   \caption{The density of fluid elements at different initial heights, where the initial heights are given in the legend, for a $\gamma = 5/3$ polytrope and $\beta = 3$ as a function of time. The different curves are for different initial cylindrical distances from the center of the star, where the top-most curve for each set of colored lines has $s_0 = 0$ (above the center of the star), while the bottom-most curve has $s_0 = 0.26$. The black points illustrate the relative maximum of each fluid element.}
   \label{fig:rho_lag_s0}
\end{figure}

Figure \ref{fig:rho_lag_s0} illustrates the density of fluid elements, relative to the initial, central density of the star, as a function of time for a $\gamma = 5/3$ polytrope and for $\beta = 3$ with $N = 4$. As for Figure \ref{fig:z_lag_s0}, the different colors correspond to the initial heights shown in the legend, and the curves within each colored bundle are for different initial cylindrical radii within the star; here the top-most curves have $s_0 = 0$ while the bottom-most curves have $s_0 = 0.26$. The black points show the relative maximum of each curve. This figure demonstrates that, for this $\beta$, the density within the compressing star is maximized for small $z_0$ and small $s_0$, i.e., the center of the star remains the location of the highest density. We also see that larger initial heights above the plane and larger initial cylindrical distances lead to the most dramatic consequences of the nonlinearity in the relation between $z$ and $z_0$. 

Figure \ref{fig:rho_lag_s0} also shows that, for an initial height of $z_0 = 1.3$, the density starts to increase again and, for initial cylindrical distances $s_0 \gtrsim 0.3$, diverges around a time of $t_{\star} \simeq 0.1$. The reason for this behavior is that, at these large heights and distances, the nonlinear effects result in the crossing of fluid elements and the formation of a shock; the crossing of the fluid elements can be seen from the blue curves in Figure \ref{fig:z_lag_s0}. Since the density is inversely proportional to the spacing between fluid elements, the crossing of two fluid elements results in a divergence of the density. We now turn to a discussion of shock formation.

\section{Shock formation and evolution}
\label{sec:shocks}
Two fluid elements initially spaced by an infinitesimal amount $\Delta z_0$ will cross when $\partial z/\partial z_0 = 0$. When $N = 1$, such that the relation between the current Lagrangian position $z$ and the initial Lagrangian height $z_0$ is given by Equation \eqref{znonhomo}, this condition occurs when

\begin{multline}
H_{00}+3H_{10} z_0^2 +H_{01}s_0^2 = 0  \\
\Rightarrow \quad z_0 = \sqrt{-\frac{H_{00}(\tau)+s_0^2H_{01}(\tau)}{3H_{10}(\tau)}}.
\end{multline}
The characteristic of the shock is therefore given by

\begin{equation}
z_{\rm sh}(\tau) = \frac{2}{3}\sqrt{-\frac{H_{00}+s_0^2H_{01}}{3H_{10}}}\left(H_{00}+s_0^2H_{01}\right). \label{Zshock}
\end{equation}

\begin{figure}[htbp] 
   \centering
   \includegraphics[width=0.475\textwidth]{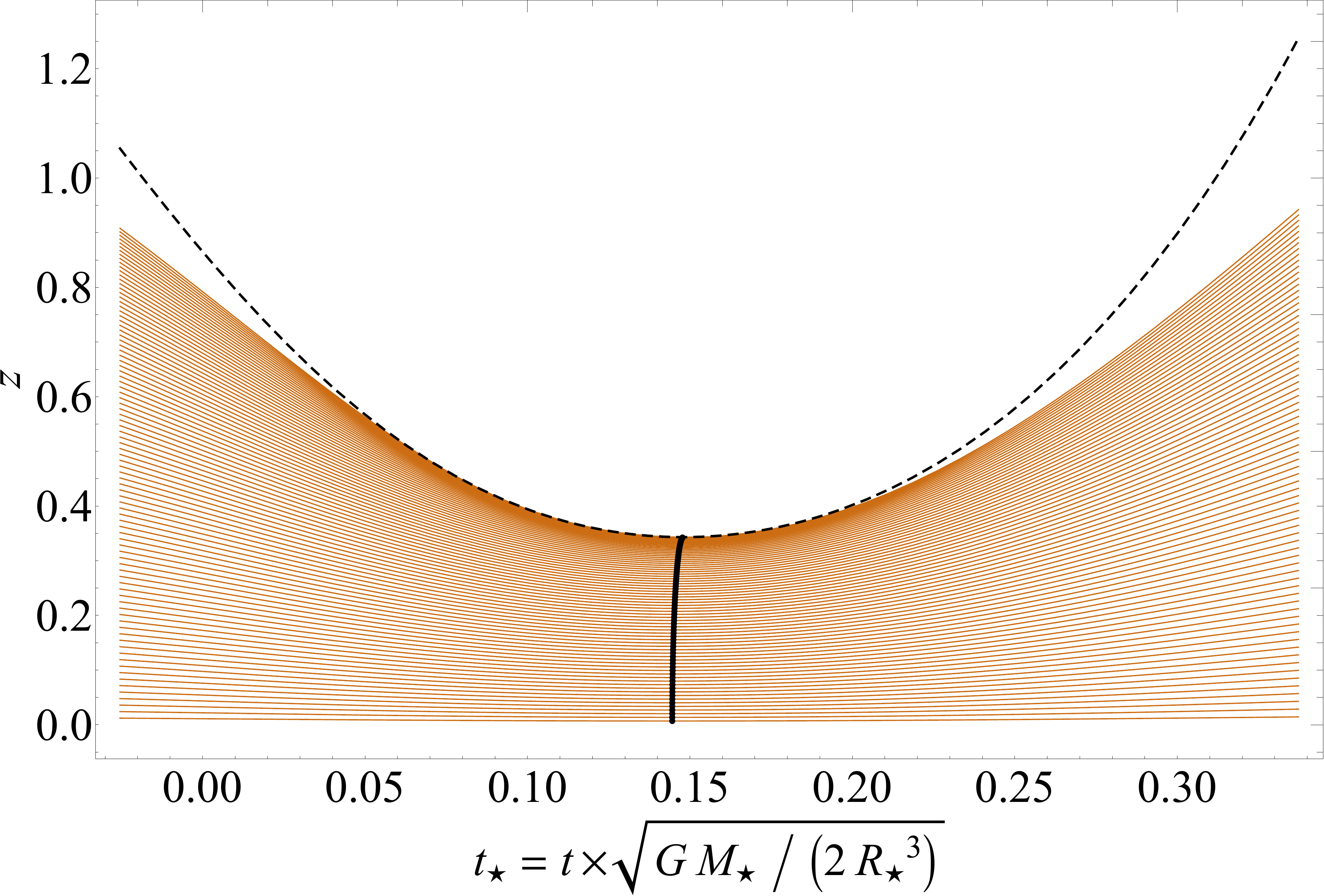} 
   \caption{The Lagrangian heights of fluid elements for the disruption of a $\gamma = 5/3$ polytrope with $\beta = 2.5$, where each orange curve corresponds to a different initial height above the plane, $z_0$, with the nonlinear solution that has $N = 1$; here the initial cylindrical distances of all the particles is set to zero ($s_0 = 0$), such that they are all directly above the center of the star, and the fluid element with the minimum initial height (bottom-most curve) has $z_0 = 0.02$ and the maximum initial height is $z_0 = 1.8$ (roughly at the surface of the polytrope). The black, dashed curve shows the location of the shock, and the points show the time of maximum compression of each fluid element. The shock initially propagates downward into the star, then stalls around the time of maximum compression, and reverses its direction of motion and moves upward.}
   \label{fig:zshock_beta2p5}
\end{figure}

The orange, solid curves in Figure \ref{fig:zshock_beta2p5} illustrate the Lagrangian heights of the fluid elements for a $\beta = 2.5$ disruption of a $\gamma = 5/3$ polytrope, where different curves correspond to different initial heights $z_0$. The bottom-most curve has $z_0 = 0.02$, while the top-most curve has $z_0 = 1.8$ (which roughly coincides with the initial surface of the star), and the points illustrate the times of maximum compression of the fluid elements. The black, dashed curve gives the location of the shock that forms within the collapsing star (when a fluid element is hit by the shock we stop showing its location for clarity) as determined from Equation \eqref{Zshock}. Initially the shock penetrates downward into the compressing gas, and starts at the surface at a time of $t_{\star} \simeq 0.07$. Interestingly, the shock then {stalls} at a time of $t_{\star} \simeq 0.15$, which is approximately coincident with when the majority of the fluid elements reach their minimum height. The shock then {reverses its direction of motion} and propagates back upward from the center of the (now rebounding) star.

The origin of this behavior of the shock is that the gas into which it is advancing is compressing and eventually rebounding and moving in a direction that opposes the motion of the shock. Thus, after the shock forms and as time progresses the stellar material beneath it is further compressed, which increases the pressure and sound speed of the gas and reduces the Mach number of the shock. The fluid beneath the shock eventually reaches maximum compression, rebounds, and then moves upward and into the advancing shock. This transfer of upward momentum causes the shock to stall and eventually reverses its motion, such that the shock does not penetrate all the way to the center of the star.

\begin{figure}[htbp] 
   \centering
   \includegraphics[width=0.475\textwidth]{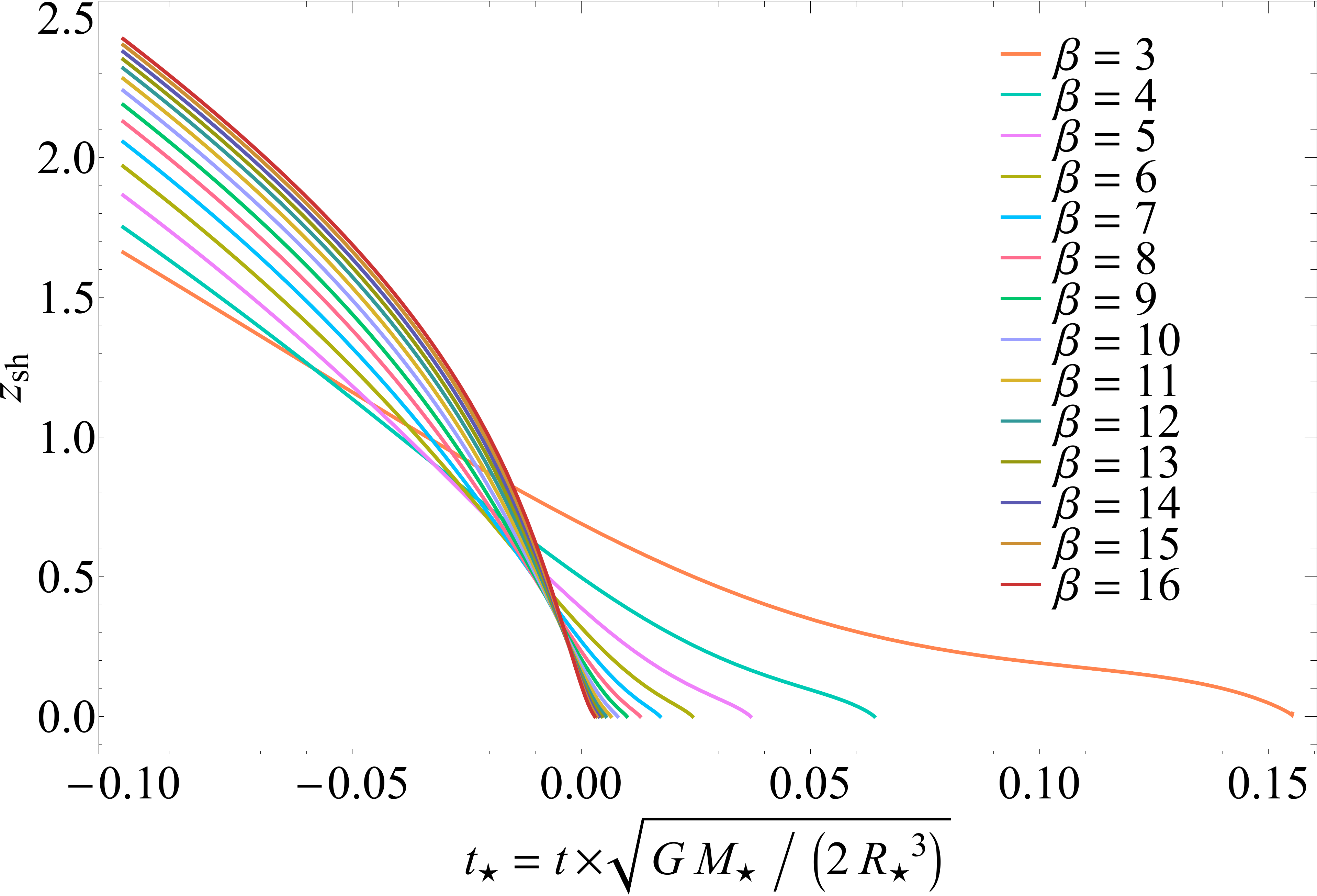} 
   \caption{The time-dependent position of the shock, as determined from Equation \eqref{Zshock}, for the $\beta$ shown in the legend. Here the initial cylindrical radius is set to zero, so that the shock propagates directly above the center of the star. Here each case reaches $z_{\rm sh} = 0$, which contrasts the shock evolution shown in Figure \ref{fig:zshock_beta2p5} where the shock propagation is halted and reversed by the compressing gas beneath it.}
   \label{fig:zsh_of_beta_1}
\end{figure}

Figure \ref{fig:zsh_of_beta_1} gives the position of the shock for $s_0 = 0$ as a function of time for the $\beta$ shown in the legend. We see that for small $\beta$ the effects of the compression of the material beneath the shock slow the propagation of the shock, but unlike the case for $\beta = 2.5$ are incapable of completely reversing its motion. As $\beta$ increases the shock is relatively unimpeded by the dynamical evolution of the gas beneath it and reaches the origin at a time that is nearly coincident with the time at which the central density of the star is maximized (i.e., at the time where maximum compression is reached for fluid elements near the midplane). The solutions for the shock position with $s_0 \neq 0$ look similar to those in this figure.

Equation \eqref{Zshock} gives the location in the fluid where two fluid elements cross with third-order terms included in the relationship between the location of a fluid element, $z$, and its initial position, $z_0$. While this measure is a good estimate of when a shock is likely to form within the flow, it is apparent that our solution for $z(z_0,\tau)$ that includes only third-order terms breaks down in regions of the flow near a shock, as this location coincides with where $\partial z_0/\partial z \rightarrow \infty$, and hence the density would increase without bound, in the limit that the number of terms in our series $N\rightarrow \infty$. Our series expansion to third order in $z_0$, on the other hand, does not diverge. Including higher-order terms in the relationship between $z$ and $z_0$ will modify the location at which our solution for the evolution of the fluid becomes highly nonlinear and where $\partial z/\partial z_0 = 0$, and are therefore essential to include when fluid elements start to cross in the lower-order solution.

Moreover, it is apparent that when a shock forms within the flow, our series expansion will {never equal the true solution beyond the shock} and will only remain accurate at $z < z_{\rm sh}$. This is because our series expansion is valid about the origin, but once $\partial z/\partial z_0$ approaches zero, the series expansion will never be able to recover the diverging nature of the pressure gradient at that point. Instead, including more terms in the series will cause the solution to more inaccurately represent the true solution for regions beyond $z_{\rm sh}$, as the series solution no longer converges beyond that point. With an ever-increasing number of terms retained in the relationship between $z$ and $z_0$, the position of the shock within the fluid therefore does not correspond to the location where fluid elements cross at some level of approximation (i.e., with a finite $N$), but is instead the distance at which our series solution no longer converges as we increase the number of terms in the relationship between $z$ and $z_0$. The shock is therefore the time-dependent radius of convergence of our series solution.

\begin{figure}[htbp] 
   \centering
   \includegraphics[width=0.475\textwidth]{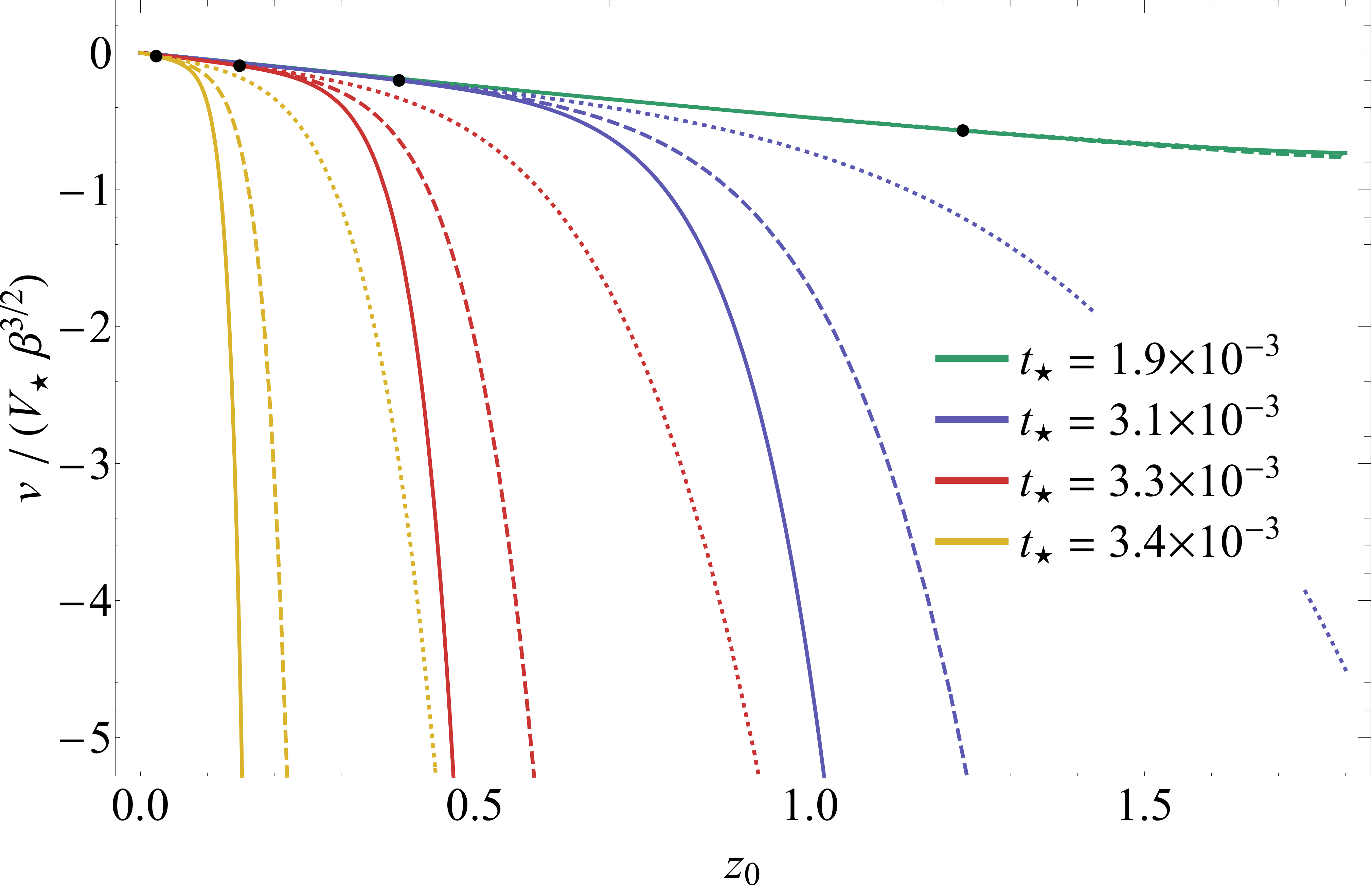} 
   \caption{The velocity $v$ normalized by $V_{\star}\beta^{3/2}$ as a function of initial position $z_0$ for a $\gamma = 5/3$ polytrope and when $\beta = 15$. The differently colored curves correspond to the times shown in the legend, and the solid curves have $N = 4$ (largest exponent in the series expansion about $z_0$ equal to 9), the dashed curves have $N = 3$ (largest exponent equal to 7), and the dotted curves have $N = 2$ (largest exponent equal to 5). The black points show the location of the shock according to our criterion that the standard deviation of the $v(z_0)$ curves exceed a threshold value.}
   \label{fig:v_of_z0_comps}
\end{figure}

Figure \ref{fig:v_of_z0_comps} gives the solution for $v(z_0,\tau)$ for $\beta = 15$ for the times shown in the legend and $N = 4$ (solid curves), 3 (dashed curves), and 2 (dotted curves), where the velocity is normalized by $V_{\star} = \alpha\sqrt{GM_{\star}/(2R_{\star}^3)}$ and $\beta^{3/2}$; recall that the maximum exponent in the expansion of $z$ about $z_0$ is $2N+1$. At the earliest time the solutions with different $N$ are nearly indistinguishable from one another over the whole range of $z_0$. However, at times of $t_{\star} = 3.1\times 10^{-3}$ and later, we see that the curves are indistinguishable up to a point $z_{\rm 0, sh}$, beyond which point they all diverge rapidly from one another. The existence of this time-dependent location in $z_0$ demonstrates that a shock has formed within the flow -- the series will not accurately capture the behavior of the true solution (which is piecewise-continuous) no matter how many terms we add. 

We define the location of the shock as the $z_0$ where the standard deviation $\sigma$ of $\partial v(z_0,\tau)/\partial z_0$ exceeds 1, where the standard deviation is defined as

\begin{equation}
\sigma^2(z_0,\tau) = \frac{1}{2}\sum_{\rm N = 4}^{6}\left(\frac{\partial v}{\partial z_0}-\mu(z_0,\tau)\right)^2
\end{equation}
and the average of the derivative of the velocity, $\mu(z_0,\tau)$, is

\begin{equation}
\mu(z_0,\tau) = \frac{1}{3}\sum_{\rm N = 4}^{6}\frac{\partial v}{\partial z_0}.
\end{equation}
We choose the derivative of the velocity to measure the standard deviation, rather than the velocity itself, because when the shock nears the origin the velocity is small everywhere even though solutions with different $N$ are discrepant. Therefore, if we used the velocity the standard deviation would be much less than one even though the solutions themselves are mutually inconsistent (with respect to their log). While our choice of the threshold value of $\sigma$ of 1 is somewhat arbitrary, the location of the shock is not sensitive to this value (as long as it is not too small) because the solutions rapidly diverge from one another -- and hence $\sigma$ quickly exceeds the threshold of $2$ -- once we are beyond the position of the shock. The black points in Figure \ref{fig:v_of_z0_comps} illustrate the position of the shock as defined by this criterion; we see that it accurately reproduces the location at which the curves diverge (note that the velocity is normalized by $\beta^{3/2} \simeq 58$ for visualization purposes in this figure, so the discrepancies between the curves are larger by a factor of $\sim 50$ than what is shown).

\begin{figure}[htbp] 
   \centering
   \includegraphics[width=0.475\textwidth]{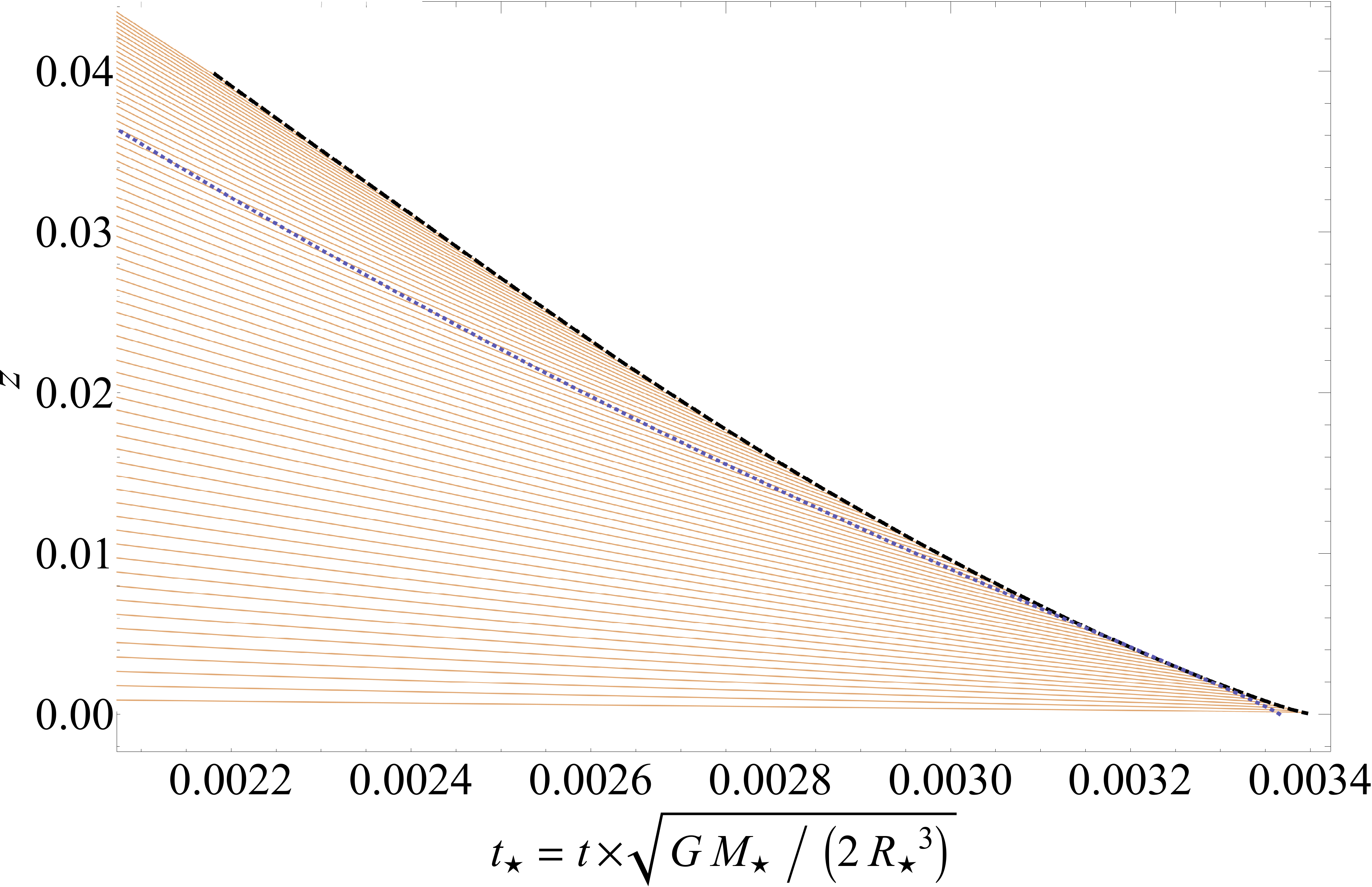} 
   \caption{The Lagrangian height above the plane $z$ for a range of initial heights $z_0$ (orange curves) as a function of time in units of the dynamical time of the star; here the star is a $\gamma = 5/3$ polytrope, the initial cylindrical distance is $s_0 = 0$, and $\beta = 15$. The black, dashed curve shows the location of the shock as defined by our criterion that the standard deviation of the derivative of the velocity as a function of $N$ exceed a critical value, and the purple, dotted curve shows the location of the shock calculated from the $N = 1$ solution when fluid elements cross (Equation \ref{Zshock}). }
   \label{fig:z_lag_shocks}
\end{figure}

Figure \ref{fig:z_lag_shocks} illustrates the Lagrangian heights of fluid elements directly above the center of mass of the star ($s_0 = 0$) for a range of initial heights (orange curves), where the smallest $z_0 = 0.02$ (lowest curve) and the largest is $z_0 = 1.4$ (uppermost curve); here the disrupted star is a $\gamma = 5/3$ polytrope and $\beta = 15$. The black, dashed curve gives the location of the shock as calculated from our criterion on the standard deviation, and the purple, dotted curve shows the location where fluid elements cross with $N = 2$ (the shock position given by Equation \ref{Zshock}). We see that the two measurements of the shock location yield very similar values; in this case the shock reaches the origin at $t_{\star} \sim 0.0033$ (recall that the stellar center of mass reaches pericenter at $t_{\star} = 0$). 

\begin{figure}[htbp] 
   \centering
 \includegraphics[width=0.475\textwidth]{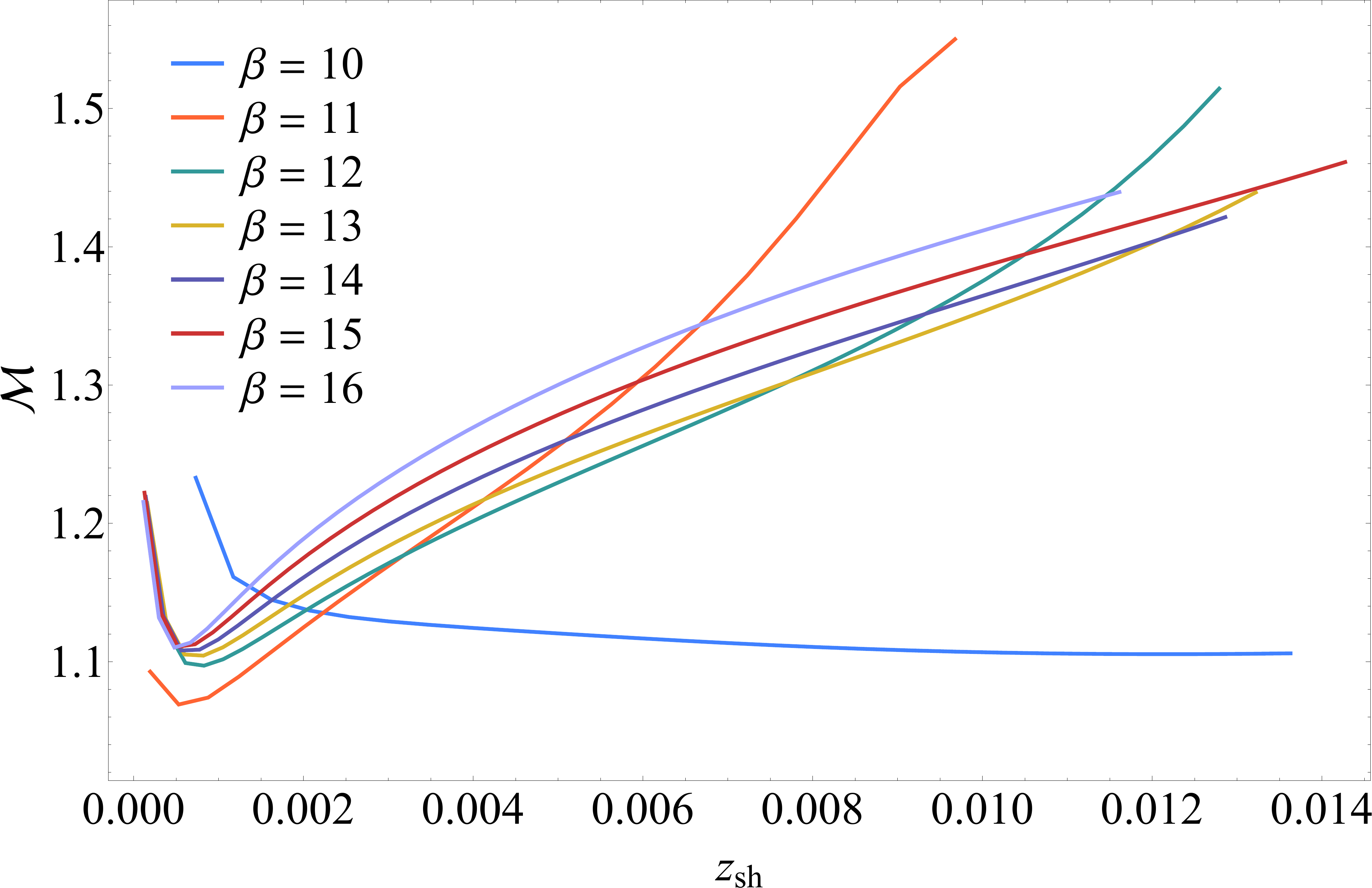}
   \caption{The Mach number of the shock as a function of the height of the shock above the plane for the $\beta$ shown in the legend (the shock propagates from large to small heights, or from right to left on the horizontal axis). The Mach number is only marginally greater than one, showing that the shock is weak; the weak nature of the shock arises because the gas beneath the shock is compressing and increasing its sound speed as the shock propagates into the collapsing star.}
   \label{fig:zsh_of_beta}
\end{figure}

Figure \ref{fig:zsh_of_beta} shows the Mach number of the shock as a function of the shock height above the plane $z_{\rm sh}$ and for the $\beta$ in the legend, where the shock Mach number is $\mathcal{M} = |v-v_{\rm sh}|/c_{\rm s}$, $c_{\rm s}^2 = \gamma p/\rho$ being the square of the sound speed of the pre-shock gas and $v$ the velocity of the pre-shock gas, and both of these quantities are evaluated immediately beneath the shock. Interestingly, this figure demonstrates that the shock is not strong, with a Mach number of $\sim 1.2$ for all $\beta$, and that the Mach number does not change substantially as the shock propagates into the compressing star. The reason that the Mach number is not large is that the gas beneath the shock is simultaneously compressing (raising its sound speed) and moving away from the shock at a substantial fraction of the shock speed. 

\begin{figure*}[htbp] 
   \centering
   \includegraphics[width=0.495\textwidth]{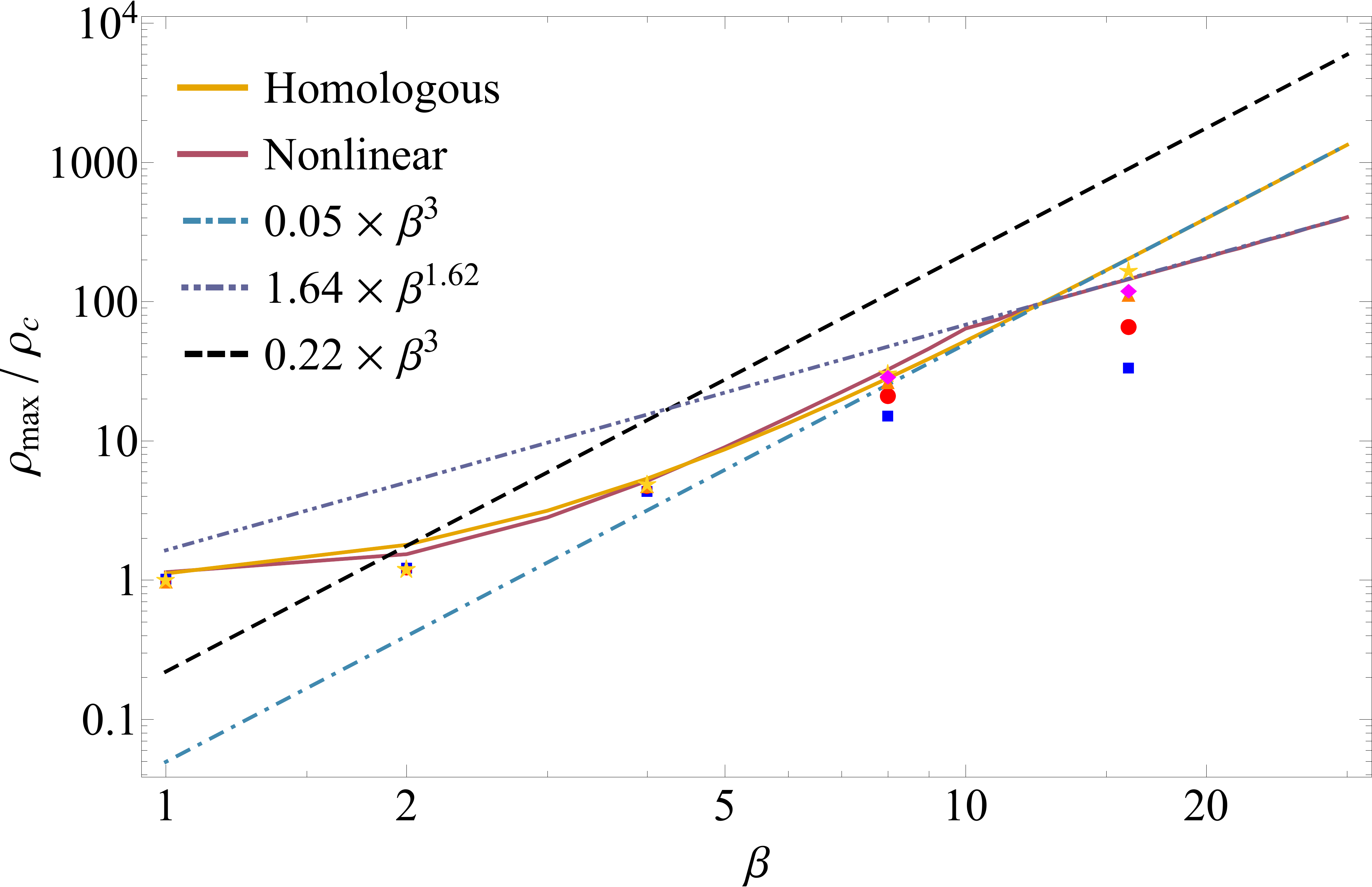} 
 \includegraphics[width=0.495\textwidth]{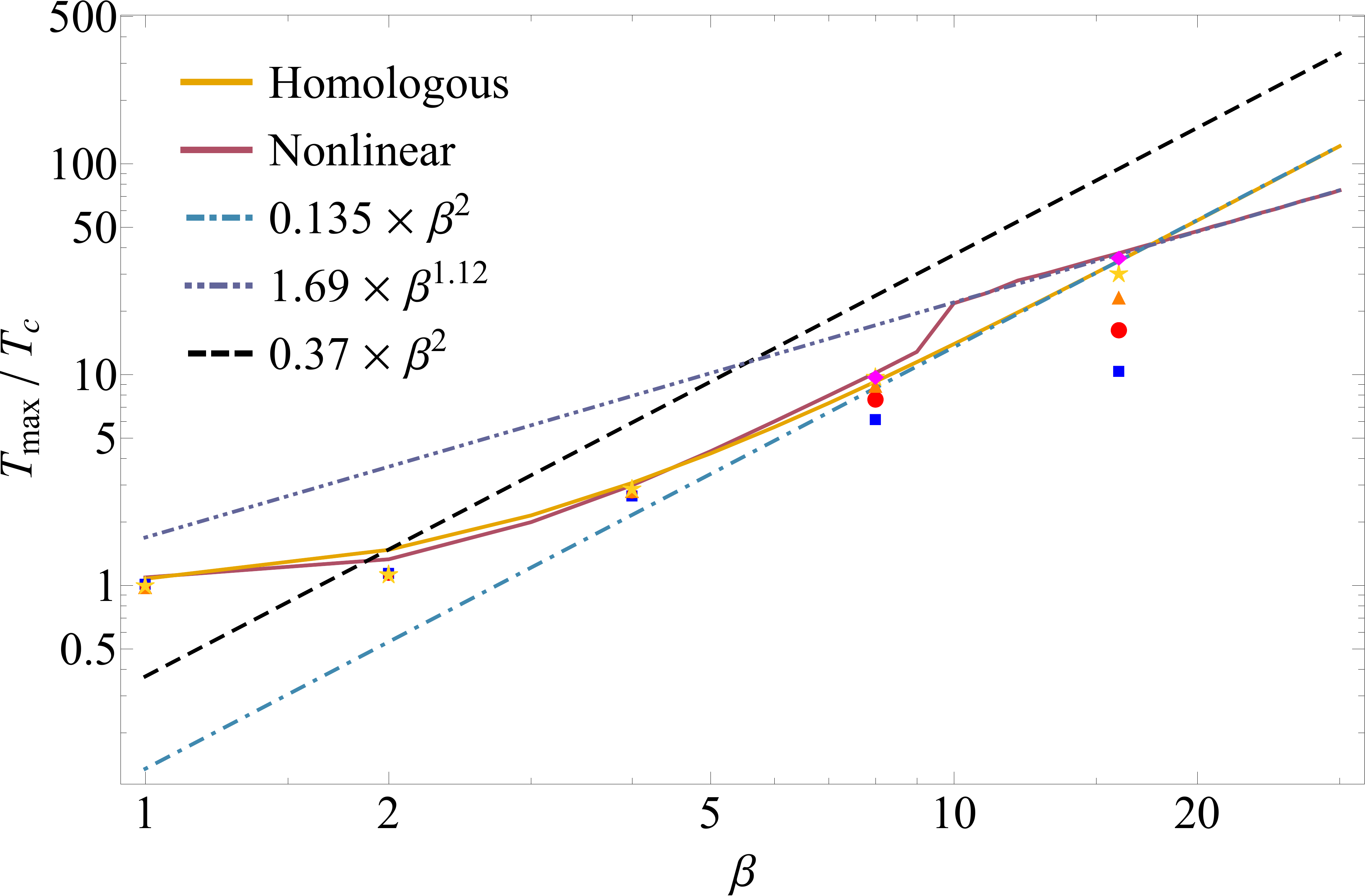} 
   \caption{The maximum density (left) and temperature (right) achieved during the compression of the star as a function of the $\beta$ of the encounter and for a $\gamma = 5/3$ polytrope; the values of the density and temperature are calculated at the geometric center of the star, or $s_0 = z_0 = 0$. The dark-yellow, solid curve shows the homologous solution, and the dark-red, solid curve is the nonlinear solution with $N = 4$ and (approximately) includes the effect that the in-plane motion of the fluid elements has on the density. The light blue, dot-dashed lines show fits to the large-$\beta$ evolution for the homologous compression, and the black, dotted line is the prediction of \citet{carter83}. Above $\beta \simeq 10$ a shock forms and reaches the origin prior to the point of maximum, adiabatic compression; the dot-dot-dashed, purple lines show fits to the large-$\beta$ limits of the shocked solutions. The points are the result of smoothed-particle hydrodynamics simulations with $0.25$ million particles (blue squares), 2 million particles (red circles), 16 million particles (orange triangles), and 128 million particles (gold stars), which were run with $\beta = 1$, 2, 4, 8, and 16. The magenta diamonds show the solution when shock heating was included in the simulation, meaning that instead of assuming that the heat generated from shocks was able to be radiated efficiently, it was retained in the specific internal energy of the fluid elements. }
   \label{fig:rho_of_t}
\end{figure*}

For $\beta \lesssim 10$, the center of the star reaches its point of maximum compression before the shock reaches the origin, and hence the adiabatic solution captures the density maximum as a function of time. For $\beta \gtrsim 10$, the maximum value of the density is given by the post-shock density as the shock reaches the origin, where the post-shock density is (from the shock jump conditions) 

\begin{equation}
\rho_{\rm ps} = \frac{\gamma+1}{\gamma-1}\left(1 +\frac{2}{\gamma-1}
\mathcal{M}^{-2}\right)^{-1}\rho. \label{rhosh}
\end{equation}
Note that this expression includes the effect of the finite Mach number, which is obviously essential to include given that, from Figure \ref{fig:zsh_of_beta}, the shock is not strong. The left panel of Figure \ref{fig:rho_of_t} shows the maximum value of the density achieved during the compression of the star as a function of $\beta$ at the center of the star ($s_0 = z_0 = 0$; when the shock reaches the midplane prior to the point of maximum compression, the maximum density is calculated at the point where the shock reaches an initial Lagrangian height of $z_0 = 0.01$). The solid, red curve results from the nonlinear solution with the in-plane stretching (approximately) included, while the yellow curve is the homologous solution. The dot-dashed, blue line is the fit to the large-$\beta$ behavior of the homologous solution. The dot-dot-dashed, purple line is a fit to the large-$\beta$ limit of the post-shock density, which we find scales as $\propto \beta^{1.62}$, which is a much weaker scaling than the adiabatic limit predicts. 

The right panel of this figure shows the maximum temperature reached normalized by the initial central temperature of the star, where the curves are analogous to the left-hand panel; again, this is for the center of the star, or $s_0 = z_0 = 0$. The post-shock temperature is calculated from the post-shock pressure, which from the jump conditions is

\begin{equation}
p_{\rm ps} = \frac{2}{\gamma+1}\left\{1-\frac{\gamma-1}{2\gamma}\mathcal{M}^{-2}\right\}\rho\left(v_{\rm sh}-v\right)^2, \label{psh}
\end{equation}
and we assumed that the gas was gas-pressure dominated so $T \propto p/\rho$. The purple, dot-dot-dashed line is a fit to the large-$\beta$ solution for the post-shock gas, and is $\propto \beta^{1.12}$; as for the density, this scaling is much shallower than the adiabatic prediction\footnote{Note that one would predict $T_{\rm max} \propto \beta^2$ for the post-shock temperature from the freefall solutions if the shock were strong, as noted by \cite{Bicknell:1983aa}; it does not follow this relationship because the shock is weak.} of $T_{\rm max} \propto \beta^2$. 

The squares, circles, triangles, and stars in Figure \ref{fig:rho_of_t} give the results of SPH simulations performed with {\sc phantom}, in which a $\gamma = 5/3$ polytrope was disrupted by a $10^6 M_{\odot}$ supermassive black hole for $\beta = 1$, 2, 4, 8, and 16 (detailed in \citealt{norman21}). The squares show the results with $0.25$ million particles, the circles with 2 million particles, the triangles with 16 million particles, and the stars with 128 million particles, and for each one of these the effects of shock heating on the gas were excluded (i.e., a shock will form in the simulation, but the heat generated therefrom is lost from the system; the density is therefore over-estimated in this case if shock heating is important). We see that the prediction from our model is in very good agreement with the highest-resolution simulations up until $\beta = 16$, at which point the simulations are not yet (convincingly) converged\footnote{As an aside, this figure also demonstrates that under-resolving the compression of the star near pericenter results in an underestimate of the density. This behavior likely explains why the simulations of \citet{Bicknell:1983aa}, which used a maximum of 2000 particles (though most were with 500) to simulate the disruption of a $\gamma = 5/3$ polytrope up to $\beta = 87$, yielded an anomalous decrease in the maximum density above $\beta \gtrsim 10$.}. The magenta triangles are simulations with 128 million particles but with the effects of shock heating included. We see that for all simulations up to $\beta = 16$, accounting for shock heating does not affect the results of the simulations, whereas for $\beta = 16$ it reduces the maximum density by a factor of roughly 1.5. This finding is in agreement with our predictions, being that the adiabatic compression can continue up until $\beta \simeq 9$, and beyond that the passage of the shock determines the maximum density within the collapsing star.

\section{{Discussion}}
\label{sec:discussion}
{Here we compare our model to the ``affine'' models in the literature, and we present a brief discussion of two other, physical effects that we have not addressed in our preceding analysis, namely the importance of the non-polytropic nature of radiative stars and general relativistic effects (see also Section 3.2 of \citealt{norman21}).}

\subsection{{Comparison to affine models}}
\label{sec:comps}
The model that we developed in Section \ref{sec:homologous} is homologous in the relationship between $z$ and $z_0$, so that the current and initial heights of fluid elements within the collapsing star are related by a function that is purely of time, and the in-plane coordinates (relative to the stellar COM) are assumed to be unchanged, i.e., $s = s_0$ for all time. The motivation for the latter approximation is that the primary contributor to changes in the density while the star is within the tidal sphere of the black hole is the vertical compression, which has been noted by other authors (though we also relaxed that assumption approximately; see Section \ref{sec:non-homologous}). Another, somewhat more concise way of writing this relationship is to say that $x^{\rm i} = T^{\rm i}_{\rm\,\,j}x_0^{\rm j}$, where in this case the tensor $T^{\rm i}_{\rm\,\, j} = \{\{H,0\},\{0,1\}\}$ and the vector of coordinates is $x^{\rm i} = \{z,s\}$ (and analogously $x_0^{\rm j} = \{z_0,s_0\}$).

The assumption of a linear relationship between the current and initial Lagrangian coordinates has been called an affine transformation, and applying such a transformation specifically to a star results in an affine-star model. These affine models have been widely used to understand the evolution of various physical systems, including tidally interacting binaries, tidal disruption events, and rotating stars (e.g., \citealt{lattimer76, carter83, carter85, kosovichev92, kochanek92, lai94, ivanov01}). The model in \citet{carter83}, as expanded upon in much more detail in \citet{carter85}, is one example of an affine model and which they used to understand the compression of the star in a deep TDE. 

Therefore, the fundamental underlying assumption about the relationship between current and initial Lagrangian coordinates between our model in Section \ref{sec:homologous} and that of any affine model is identical\footnote{{That the isodensity contours of an affine star are ellipsoids is a ramification of this more fundamental assumption, and not an additional assumption itself, is evident from inserting the $r_0(r)$ relationship into the expression for the leading-order density profile of the star, $\rho \propto 1-r_0^2$.}}. The differences between the two models arise from the method employed for determining the temporal evolution of the coefficients that enter the connection. Affine-star models have used global conservation conditions, and use integrated quantities over the entire volume of the star, to constrain the evolution of the time-dependent matrix coefficients. Here, on the other hand, we use the fluid equations in differential form and equate leading-order terms in the initial Lagrangian coordinates, as the fluid equations must hold identically for any choice of $z_0$.

As we have made clear, the homologous (or affine) relationship is only correct to leading order in the initial Lagrangian coordinates, a direct corollary of which is that only the quadratic dependence of the density and pressure on the coordinates (valid near the center of the star) is accurately captured by the homologous solution\footnote{{The fact that the affine relationship is the leading-order term in a series expansion was also appreciated by \citet{carter83}, as they remark, ``In order to interpret this model in relation to a more accurately realistic hydrodynamic description in terms of the full system of field equations, the linear relation is to be thought of as the first term in a power series expansion for the actual value of the position coordinates $r_{\rm i}$ (relative to the centre) as functions of their original values $\hat{r}_{\rm i}$.'' Whether or not they realized that this leading-order nature of the affine model also restricted the accuracy of the solution in terms of the density profile of the star is less clear.}}. It therefore seems unreasonable, and we argue not self-consistent, to enforce global conditions related to and including the surface of the star on a model that only accounts for the density variation deep within the stellar interior (see Figure 2 of \citealt{coughlin21b} for the number of terms necessary in a series expansion to accurately approximate the surface features of a polytrope). We find that our method of equating the terms that appear in differential form in the fluid equations, which relate fluid phenomena locally, more accurately reflects the restrictive nature (in space) of the affine approximation.

Another consequence of this restricted applicability, in a somewhat different context, is that the solutions of either the homologous model we presented here or the affine models studied elsewhere -- both of which incorporate nonlinear effects -- will not accurately reproduce the linear behavior of an oscillating star as computed from linear perturbation theory. The reason for this is, as we have already stated, that the affine relationship between the current and initial Lagrangian fluid elements is only accurate in the deep interior of the star, and therefore does not account for variations in the fluid properties near the stellar surface, whereas the eigenvalues calculated from linear perturbation theory do account for this region of the star (but at the expense of dropping nonlinear terms). In fact, it was shown by \citet{coughlin21b} that accurately reproducing the eigenvalues themselves and the number of eigenvalues that characterize the radial oscillations of a star necessitates going to higher order (i.e., accounting for the nonlinear relationship between the current and initial Lagrangian coordinates), and it is not surprising that the ellipsoidal (affine) models of \citet{kochanek92, kosovichev92, lai94} did not reproduce the linear tidal theory (see, e.g., Section 7.3 of \citealt{lai94}). A simple way to see this is to realize that by having a homologous relationship between the coordinates we can only have a homologous velocity profile, i.e., $v \propto z$, and there are no zero crossings in the velocity (the velocity can only be zero everywhere at a given time). The higher-order modes, however, contain zero crossings for all eigenmodes greater than the first. It therefore follows that the number of eigenvalues that we can capture through this approach is equal to the number of terms in the expansion between $z$ and $z_0$ (or, for a spherically symmetric star as in \citealt{coughlin21b}, the current and initial Lagrangian shells $r$ and $r_0$). 

The reason that the lower-order modes are more accurately reproduced when going to higher order is that the equations describing the coefficients are coupled, and therefore the dynamical equation derived describing (for our case) $H(\tau)$ at the homologous level is not the same as the one that accounts for the nonlinear terms; we illustrated this behavior directly in Section \ref{sec:non-homologous}. Finally, we emphasize that while our numerical results shown above (Figure \ref{fig:rho_of_t}) provide verification of our model in the context of tidal disruption events (and demonstrate disagreement with the affine model of \citealt{carter83} and the numerical results of \citealt{brassart08}), the results of \citet{coughlin21b} provide distinct verification of the validity of this method by demonstrating that it produces both the eigenvalues and the nonlinear mode couplings of a star once sufficiently high order is reached.

\subsection{Non-polytropic stars}
\label{sec:non-polytropic}
\begin{figure}[htbp] 
   \centering
   \includegraphics[width=0.475\textwidth]{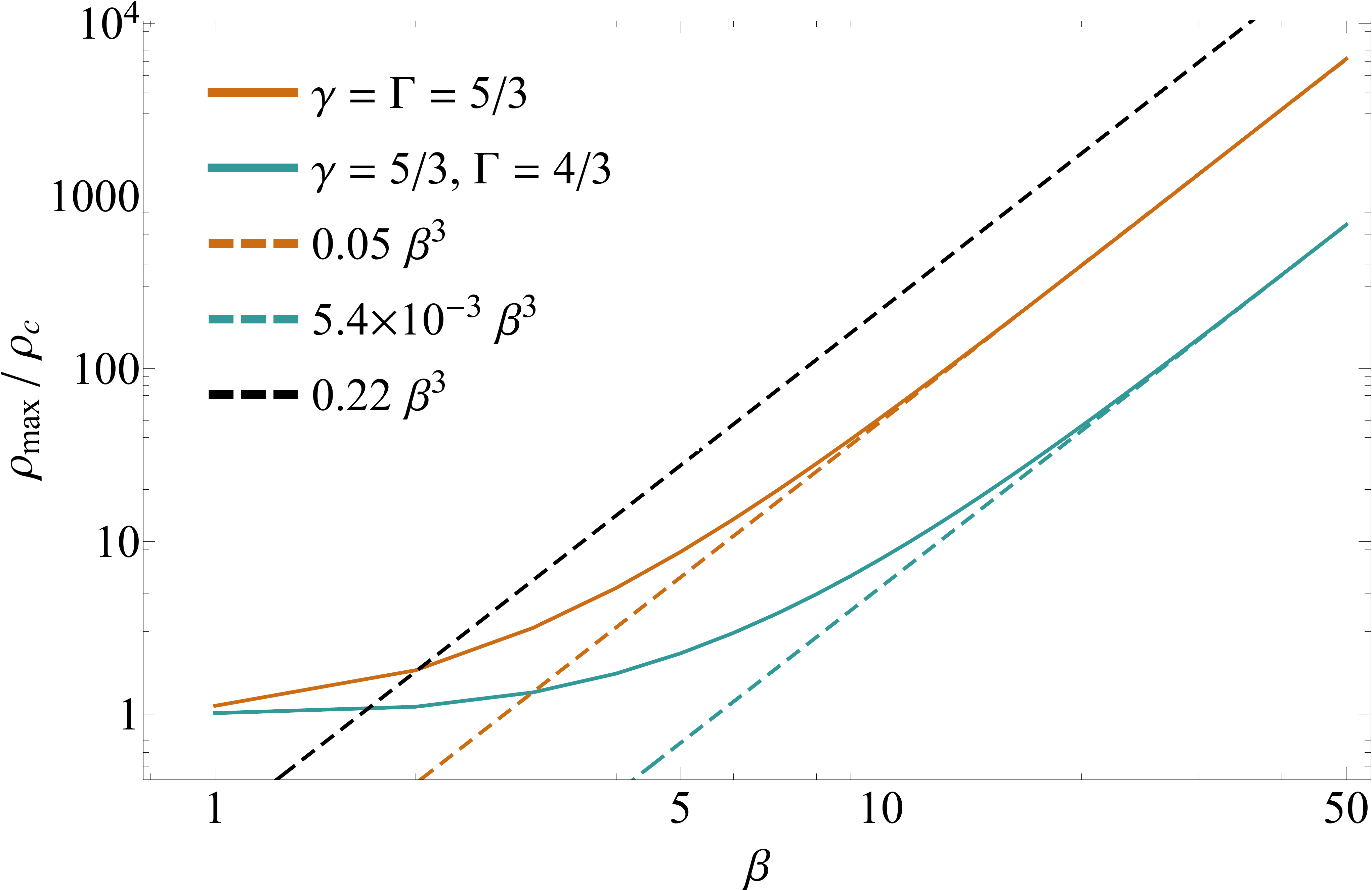} 
   \caption{The maximum density as a function of $\beta$ for a $\gamma = 5/3$ polytrope (orange, solid line) and a $\Gamma = 4/3$ polytrope with an adiabatic $\gamma = 5/3$ (blue, solid line). The orange-dashed and blue-dashed lines are fits to the large-$\beta$ behavior of the solution, while the black, dashed line is the prediction from \citet{luminet86}.}
   \label{fig:rhomax_np}
\end{figure}

So far we assumed that the star being disrupted was polytropic. While this is a good approximation for low-mass and fully convective stars, radiative stars are better approximated by $\Gamma = 4/3$ polytropes with $\gamma \simeq 5/3$ (the Eddington standard model; \citealt{hansen04}). In this case it is simple to demonstrate that the equation describing the homologous evolution is precisely the same as that derived in Section \ref{sec:homologous}, namely Equation \eqref{Hdyn}, but with $\rho_{\rm c}/\rho_{\star} \simeq 54.2$ (appropriate to a $\Gamma = 4/3$ polytrope). Figure \ref{fig:rhomax_np} gives the maximum density as a function of $\beta$ for a star described by the Eddington standard model ($\Gamma = 4/3$ and $\gamma = 5/3$), as shown by the solid, blue line. For reference we have also shown the result for a $\Gamma = \gamma = 5/3$ polytrope (solid, orange line). The $\propto \beta^3$ relation starts to hold once $\beta \gtrsim 20$, which is a result of the fact that the large central pressure is able to withstand the tidal compression much more effectively. On this same plot we also show the prediction of \citet{luminet86} for a $\gamma = 5/3$ polytrope for comparison. 

We could perform all of the same analysis in Section \ref{sec:non-homologous} to work out the nonlinear contributions to the equations motion for the gas parcels in the collapsing star and elucidate the effects that the non-zero entropy gradient have on the compression and shock formation. We leave such work to a future investigation.

\subsection{{General relativity}}
\label{sec:gr}
The analytical model that we developed, and the simulations to which we compared the predictions of the model (see \citealt{norman21}), treated the gravitational field of the supermassive black hole as a Newtonian point mass. The reasons for doing so include relative simplicity -- the analytical formalism we developed in Sections \ref{sec:homologous} and \ref{sec:non-homologous} is itself novel (as far as we are aware) and it seems reasonable to delimit the implications of this formalism in what is a relatively simple set of assumptions -- and historical precedent. Regarding the latter, the original works that described the implications of extreme compression in tidal disruption events were at least implicitly assuming a Newtonian background, and the most direct comparison of our work to theirs is enabled by adopting that same assumption.

We emphasize, however, that relativistic effects can become important when the gravitational radius of the black hole is a sizable fraction of the pericenter distance of the stellar center of mass; specifically, we have

\begin{equation}
\frac{r_{\rm p}}{r_{\rm g}} = 4.7\left(\frac{\beta}{10}\right)^{-1}\left(\frac{R_{\star}}{R_{\odot}}\right)\left(\frac{M_{\star}}{M_{\odot}}\right)^{-1}\left(\frac{M_{\bullet}/M_{\star}}{10^{6}}\right)^{-2/3}
\label{rtrg},
\end{equation}
or, adopting a radius-mass relationship of $R_{\star}/R_{\odot} = 1.06 \left(M_{\star}/M_{\odot}\right)^{0.945}$ for stars on the zero-age main sequence with masses $M_{\star}/M_{\odot} \le 1.66$ \citep{demircan91}, 

\begin{equation}
\frac{r_{\rm p}}{r_{\rm g}} \simeq 5.0\left(\frac{\beta}{10}\right)^{-1}\left(\frac{M_{\star}}{M_{\odot}}\right)^{-0.055}\left(\frac{M_{\bullet}/M_{\star}}{10^{6}}\right)^{-2/3}.
\end{equation}
Setting $r_{\rm p} = 4 r_{\rm g}$ for direct capture (see \citealt{kesden12} for the generalization to a Kerr black hole), this expression illustrates that the $\beta$ of the encounter must satisfy

\begin{equation}
\beta < \beta_{\rm DC} = 12.5\left(\frac{M_{\bullet}/M_{\star}}{10^{6}}\right)^{-2/3} \label{betaeq}
\end{equation}
to avoid direct capture (here we ignored the very weak, additional dependence on the stellar mass that scales as $M_{\star}^{-0.055}$). This equation shows that for a star with solar-like properties, for $\beta  \gtrsim 10$ the black hole will directly capture the star, as the zero-energy orbit for a Schwarzschild black hole -- within which the star (or at least the center of mass) is captured by the black hole -- is $4 r_{\rm g}$; see \citealt{kesden12} for the generalization to a Kerr black hole. For even larger $\beta$, larger black holes, or more compact stars, relativistic effects become even more important for an even larger region of the parameter space.
Indeed, for $M_{\bullet}/M_{\star} \gtrsim 10^{8}$, a solar-like star cannot be successfully destroyed by tides outside of the horizon of the hole (i.e., $r_{\rm t}/r_{\rm g} = 1$ for $\beta = 1$; \citealt{kesden12, stone19}). Most TDEs have been inferred to be hosted by black holes with masses in the range $\sim 10^{6-7}M_{\odot}$ \citep{mockler19}, for which the encounter cannot achieve a large $\beta$ before the interaction leads to a direct capture, e.g., $\beta = 2.7$ or larger results in a direct capture when $M_{\bullet} = 10^{7} M_{\odot}$. 

On the other hand, Equation \eqref{rtrg} implies that reducing the mass of the black hole yields a broader range of $\beta$ over which the Newtonian approximation is better upheld; for $M_{\bullet}/M_{\star} = 10^{5}$, for example, the prefactor in front of Equation \eqref{rtrg} is increased to $\sim 21.9$. Nonetheless, relativistic effects will still modify the tidal field and strengthen the degree to which the star is tidally compressed and stretched, even for low-mass black holes. While a detailed analysis of the impact of relativity on the tidal stress is outside the scope of our work here, a very simple but approximate estimate (though we acknowledge that this estimate has not been verified by more accurate tests; see also \citealt{tejeda13} for a comparison of other pseudo-Newtonian potentials to a relativistic treatment) of the relative increase in the strength of the tides can be found by using the Paczy\'nski-Wiita potential \citep{paczynski80} in place of the Newtonian one:

\begin{equation}
\Phi_{\bullet} \rightarrow -\frac{GM_{\bullet}}{r-2r_{\rm g}} = \Phi_{\rm N}\left(1-\frac{2GM_{\bullet}}{rc^2}\right)^{-1}, \label{Phipw}
\end{equation}
where $\Phi_{\rm N} = -GM_{\bullet}/r$ is the Newtonian expression for the potential. While clearly an approximation, this potential does reproduce salient features of the exact, general relativistic solution, including the locations of the zero-energy and innermost stable circular orbits, and it suggests that the strength of the tidal field of the black hole increases over the non-relativistic one by a factor of $(1-2r_{\rm g}/r_{\rm p})^{-1}$. For example, for a $10^{6}M_{\bullet}$ black hole and $\beta = 5$, Equation \eqref{Phipw} suggests that the tides due to general relativistic gravity are a factor of $\sim 12\%$ larger than in the Newtonian approximation. 

\citet{gafton19} recently investigated the disruption of a $\gamma = 5/3$ polytrope by a $10^6M_{\odot}$ supermassive black hole, and compared the degrees of compression suffered by the star in the Newtonian and relativistic regimes (fixed background Kerr metric). The right panel of their Figure 1 illustrates the relative importance of general relativistic effects as the $\beta$ of the encounter increases: as the pericenter distance is reduced (i.e., as $\beta$ increases), the difference between the  relativistic and Newtonian estimates of the maximum in the central stellar density increases. For the largest $\beta$ that those authors considered ($\beta = 11$, which corresponds to a pericenter distance of $4.29\, r_{\rm g}$), the difference between the Newtonian and relativistic results were less than a factor of 2, and for a non-rotating (Schwarzschild) or retrograde black hole were to within a factor of 1.5.

Redoing the analysis presented here in a general relativistic framework is outside the scope of the present work, one of the main aims of which was to compare to the (Newtonian) estimates extant in the literature. We defer an extension of the method outlined here to a relativistic one to future work.

\section{Summary and Conclusions}
\label{sec:summary}
We analyzed the tidal disruption of a star by a supermassive black hole when the pericenter distance of the star is both comparable to and well within the tidal radius of the black hole. We first briefly considered the limit in which the fluid elements comprising the star are assumed to freefall in the tidal field of the black hole (Section \ref{sec:pressureless}). While this limit has been investigated before, here we demonstrated that the pressure gradient becomes comparable to the tidal force at a time significantly before the gas pressure equals the ram pressure of the material in the freefall limit. We then developed a model that relates the current height, $z$, and initial height, $z_0$, of a Lagrangian fluid element through a homologous function of time $H(t)$, i.e., $z = H(t) z_0$. While such a relationship is exact when the gas is in pressureless freefall, we showed that it is self-consistent to leading order in $z_0$ when the effects of pressure and self-gravity are included and that the function $H$ obeys a dynamical equation that includes these effects (Section \ref{sec:homologous}). Because the dynamical equation includes both pressure and self-gravity, the model obviates the need for the impulse approximation, which assumes that the star retains perfect hydrostatic balance up until the tidal radius, and instead adopts the physically appropriate initial condition of hydrostatic equilibrium when the star is infinitely far from the black hole. 

\begin{figure}[htbp] 
   \centering
   \includegraphics[width=0.475\textwidth]{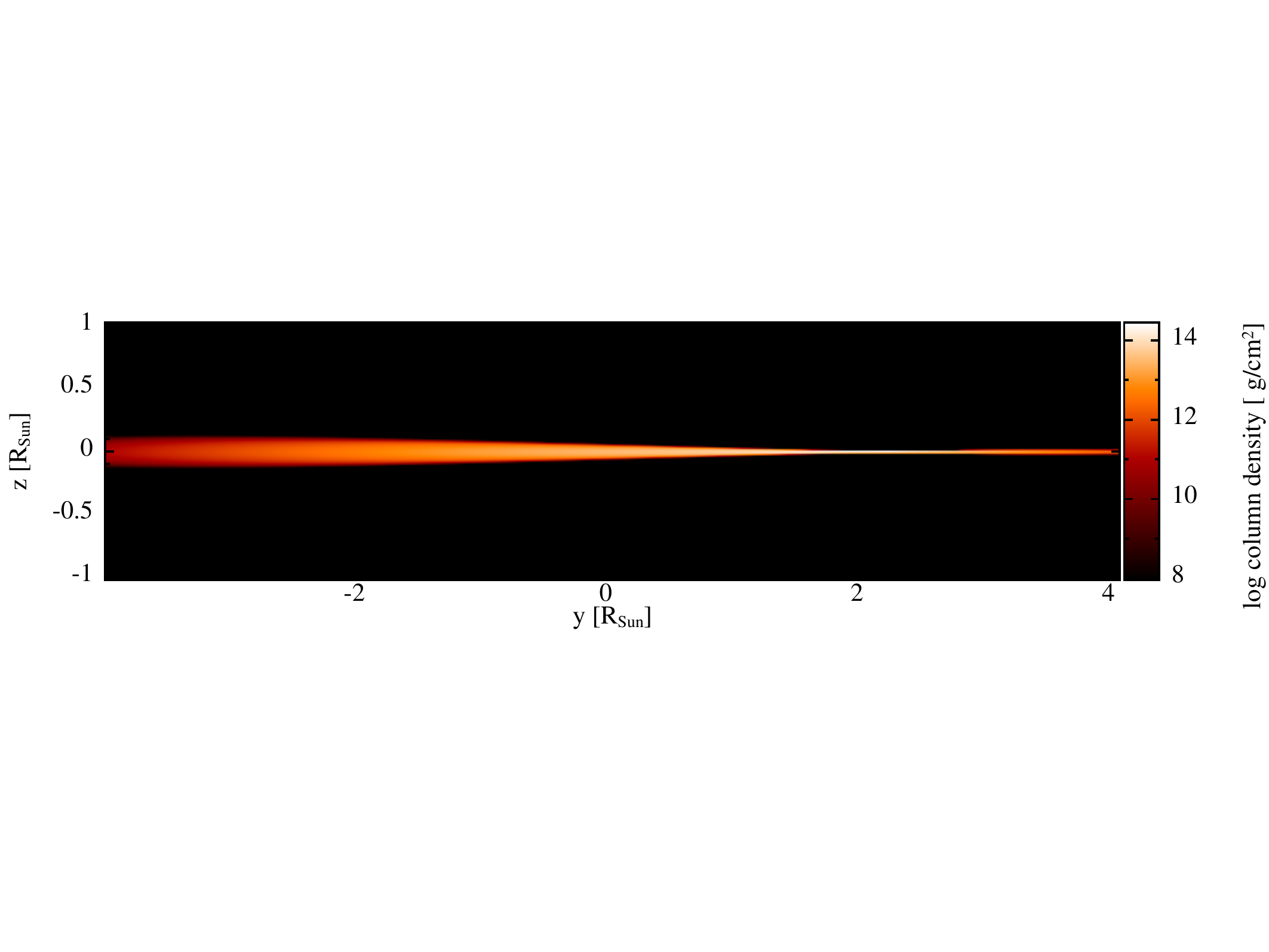} 
   \includegraphics[width=0.475\textwidth]{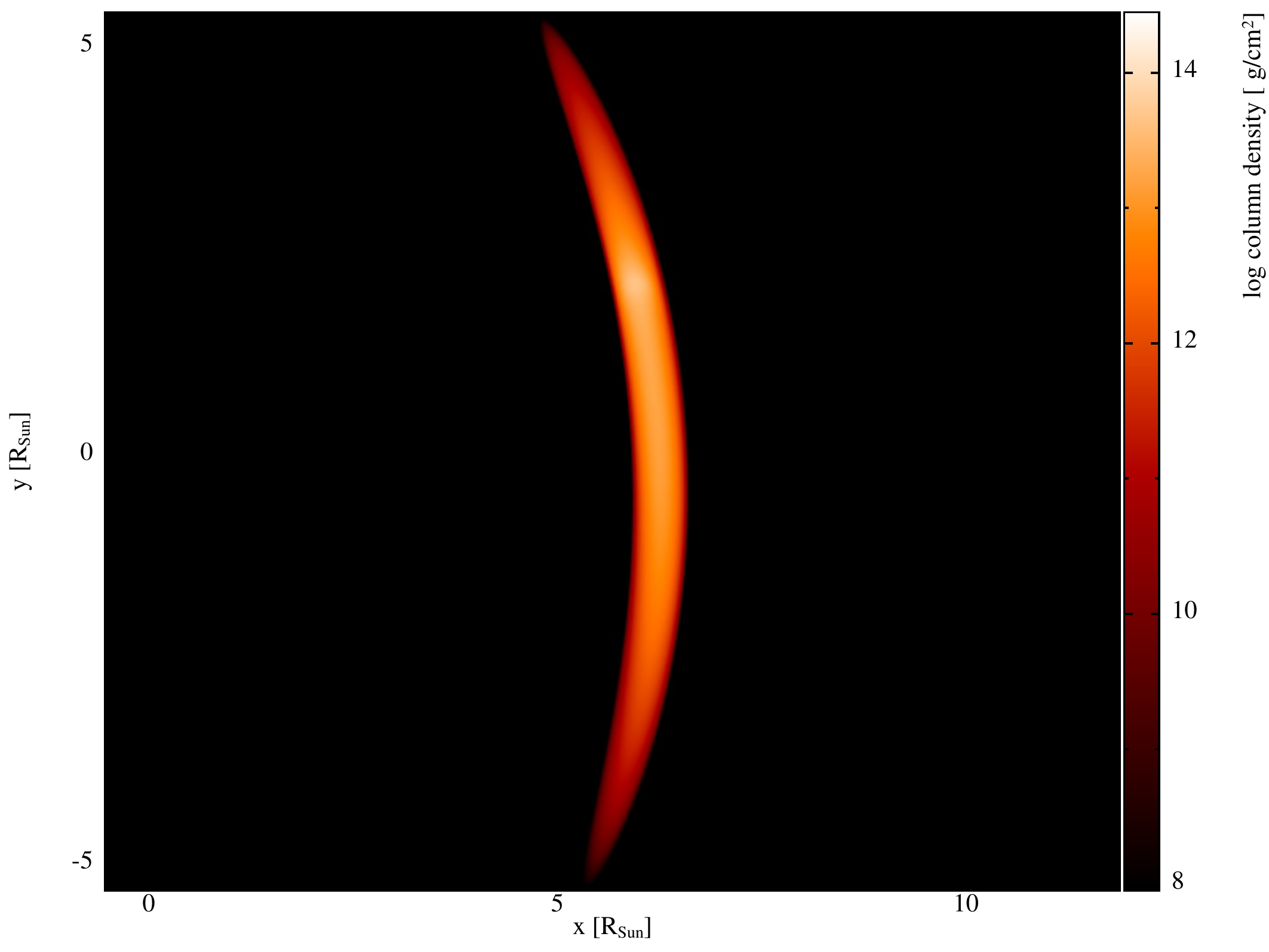} 
   \caption{The column density across the orbital plane (top) and through the orbital plane (bottom) of the original star for the $\beta = 16$, 128M-particle SPH simulation performed with {\sc phantom} (see also \citealt{norman21}); here $x$ is in the direction of the pericenter of the center of mass. At this time the center of mass of the star is at pericenter, and bright (dim) regions show increased (reduced) density.  }
   \label{fig:beta16_sim}
\end{figure}

Using this model, we demonstrated that the density at the point of maximum compression in a deep tidal disruption event does not obey $\rho_{\rm max} \propto \beta^3$ for a $\gamma = 5/3$ polytrope until $\beta \gtrsim 10$, and the coefficient of proportionality multiplying the scaling is a factor of $\sim 5-10$ smaller than that predicted in previous studies. This result is consistent with our physical argument that the pressure gradient resists the tidal compression of the black hole when the pressure is only a fraction of the freefalling ram pressure. {However, as $\beta$ increases the compression still becomes extreme: for $\beta = 16$ the maximum central density achieved during the tidal encounter is $\gtrsim 100$ times the original, central stellar density, corresponding to a decrease in the height of the star to $\lesssim 1\%$ its original value; see Figure \ref{fig:H_Hff}, \ref{fig:H_of_t_lambda10_g53} and \ref{fig:rho_of_t} (left panel). The top panel of Figure \ref{fig:beta16_sim} shows the integrated column density out of the orbital plane of the stellar center of mass (i.e., the orbital plane is the $x$-$y$ plane, and this figure shows the $y$-$z$ plane; $x$ is in the direction of the pericenter of the center of mass orbit) for the $\beta = 16$, 128M-particle disruption performed with {\sc phantom} (see \citealt{norman21} for more details of the specifics of the simulation) at the time that the original stellar center of mass reaches pericenter, while the bottom panel shows the projection onto the $x$-$y$ plane. It is apparent that the star has been flattened, or crushed, into a small fraction of the original volume it comprised. }

We then extended our model to incorporate the nonlinear terms that arise in the relationship between the current height $z$ and the initial height $z_0$ and the cylindrical radius $s_0$ as a consequence of pressure and self-gravity (Section \ref{sec:non-homologous}). We compared the results of our model to $\sim 128$ million-particle SPH simulations, finding excellent agreement with the motion of fluid elements (Figure \ref{fig:lag_simulation_comps}) and -- especially after approximately accounting for the in-plane motion of the fluid elements -- the increase in the density as the star approaches its pericenter (Figure \ref{fig:rho_comps}). 

A second consequence of the nonlinear relationship between $z$ and $z_0$ is that a shock can form in the collapsing flow, which we analyzed in Section \ref{sec:shocks}. We demonstrated that a shock forms during the compression of the star once $\beta \gtrsim 3$, and that it propagates to the origin before the gas reaches its point of maximum adiabatic compression once $\beta \gtrsim 10$. Because of the fact that the fluid beneath the inward-propagating shock is compressing, the sound speed of the pre-shock gas is increasing and the gas into which the shock is advancing is receding from and, in cases where maximum adiabatic compression is reached, running into the shock. For $\beta = 2.5$, we found that the rebound of the pre-shock gas and the transfer of upward momentum into the shock causes the shock to stall and reverse its direction of motion. The Mach number of the shock is also of the order unity, and hence the shock is not strong, over a wide range of $\beta$ (Figure \ref{fig:zsh_of_beta}) and even as $\beta$ becomes large.

Figure \ref{fig:rhomax_np} shows the maximum density (left) and temperature (right) achieved by the compressing gas during a TDE as a function of $\beta$ both from our model and high-resolution simulations. In addition to demonstrating excellent agreement between the two approaches, this figure also illustrates that shocks do not have a large effect on the TDE until $\beta \gtrsim 10$. Beyond roughly this value of the impact parameter, the shock propagates into the star and reaches the midplane before the center of the star compresses maximally and adiabatically, which reduces the maximum-achievable density below the $\propto \beta^3$ scaling predicted by \citet{carter83}. Our results therefore agree with \citet{Bicknell:1983aa}, who argued that shocks would prematurely halt the compression of the star. However, our simulation results also show that the simulations of \citet{Bicknell:1983aa} were underresolved (our highest-resolution simulations at 128 million particles are still not completely converged\footnote{This also shows that simulating $\beta \gtrsim 4$ requires at least several million particles to reliably constrain the maximum density achieved during the compression of the star, while going to $\beta \gtrsim 15$ requires at least $\sim 10^{9}$ particles, which is currently infeasible.} at $\beta = 16$ in the maximum density achieved; \citet{Bicknell:1983aa} used 500 or 2000 particles to model $\sim$ the same encounter). Beyond $\beta \simeq 10$, we find that the maximum density achieved during compression is $\rho_{\rm max}/\rho_{\rm c} \propto \beta^{1.62}$, which is a substantially weaker scaling with density than the adiabatic prediction $\rho \propto \beta^3$. Similarly, for $\beta \gtrsim 10$ we find that the maximum temperature achieved is $T_{\rm max} \propto \beta^{1.12}$; this nearly linear increase with $\beta$ is also much shallower than the expectation from adiabatic compression of $T_{\rm max} \propto \beta^2$. Therefore, the scaling of the maximum density and temperature never follow $\propto \beta^3$ and $\propto \beta^2$ for a $\gamma = 5/3$ polytrope: the scalings are much shallower for $\beta \lesssim 10$ because of the importance of the pressure gradient at early times, and they are much shallower for $\beta \gtrsim 10$ because of the effects of shock formation.

While the results of our model and three-dimensional SPH simulations are mutually self-consistent, they are inconsistent with the findings of \citet{brassart08}, who performed one-dimensional hydrodynamical simulations of TDEs (with $\gamma = 5/3$ and $\gamma = 4/3$ polytropes) up to $\beta = 15$ and found good agreement with the $\rho_{\rm max} = 0.22 \beta^3$ prediction of \citet{luminet86} and that shocks did not form during the compression of the star until $\beta \simeq 12$. Even at the homologous level of our model, we find that the maximum density is below the prediction of \citet{luminet86} by a factor of 5 -- 10, and agrees with our assessment that the pressure gradient is dynamically important and able to resist further compression when the pressure is only a fraction of the freefall ram pressure. 

We (very) briefly analyzed the case where the star being disrupted is non-polytropic (Section \ref{sec:non-polytropic}), and specified the analysis to the situation where the star is a $\Gamma = 4/3$ polytrope ($p \propto \rho^{4/3}$) and is gas-pressure dominated with $\gamma = 5/3$ (the Eddington standard model). Because of the increased central density, the maximum density achieved during the adiabatic compression does not follow the $\propto \beta^3$ scaling until $\beta \gtrsim 15$ (Figure \ref{fig:rhomax_np}). We leave a more detailed and thorough investigation of shock formation and non-homologous evolution of the fluid to future work.

\citet{Bicknell:1983aa} concluded that very little energy would be released by the triple-$\alpha$ process as a byproduct of the tidal compression. Our results -- both the model presented here and the three-dimensional SPH simulations -- yield qualitatively similar conclusions. For Helium burning to occur we require a core temperature of at least $\sim 1.2\times 10^{8}$ K \citep{hansen04}, which necessitates $\beta \gtrsim 10$ from the right-hand side of Figure \ref{fig:rho_of_t} if we assume that the initial star had a central temperature of $\sim 10^{7}$ K. As argued by \citet{Bicknell:1983aa}, the time spent at such large temperatures is very short, and not much energy would be generated from nuclear burning. The fact that a more realistic stellar progenitor that obeys the Eddington standard model suffers an even shallower increase in the central density with $\beta$ (Figure \ref{fig:rhomax_np}) implies that for solar-like star the augmented nuclear burning rate would be even smaller until $\beta$ is very large ($\gtrsim 20$). We leave a more thorough analysis of the possibility of nuclear detonation in these very high-$\beta$ encounters to a future investigation.

In Section \ref{sec:gr} we included a discussion of the effects of general relativity in the context of the applicability of our Newtonian models (both the analytical model developed here and the numerical simulations to which we made comparisons; see also Section 3.2 of \citealt{norman21}). As the black hole mass increases, general relativistic effects become more important at more modest $\beta$ owing to the fact that $r_{\rm t} = R_{\star}\left(M_{\bullet}/M_{\star}\right)^{1/3}$ while $r_{\rm g} = GM_{\bullet}/c^2$, i.e., the gravitational radius scales more strongly with black hole mass than does the tidal radius for fixed stellar properties. {}{Two important consequences of relativistic gravity are: 1) for a portion of the parameter space studied here and in previous analyses in the Newtonian approximation, the star can be captured by the black hole (i.e., have a pericenter distance $\lesssim 4 r_{\rm g}$, the zero-energy orbit) or enter its Schwarzschild (or equivalent for Kerr) radius, making the disruption unobservable; this issue is relevant for black holes with masses $M_{\bullet} \gtrsim 10^{6}M_{\odot}$, for which the direct capture radius occurs at $\beta \simeq 12.5$ for a solar-like star (i.e., one with a solar mass and radius), and is only exacerbated as the black hole mass increases -- for black hole masses $M_{\bullet} = 10^{7} M_{\odot}$ at $\beta \gtrsim 2.7$ the star is directly captured if it is solar-like (e.g., \citealt{kesden12, stone19}; see Equations \ref{rtrg} -- \ref{betaeq}); and 2) even for lower-mass black holes where the ratio $r_{\rm t}/r_{\rm g}$ is on the order of tens for $\beta \gtrsim 10$, the stresses induced by the general relativistic gravitational field can be tens of percent larger than the Newtonian value, as suggested by the simple estimate in Section \ref{sec:gr}, and can be much larger when the star nears the direct capture radius; for a $10^6M_{\odot}$ black hole, this is at $\beta \simeq 12.5$ for a solar-like star and $\beta \simeq 2.7$ for a $10^7 M_{\odot}$ black hole (and, again, a solar-like star; see Equation \ref{betaeq}). Conversely, when the Newtonian approximation is accurate (for lower-mass black holes or larger stars), our results demonstrate that the entire tidal interaction occurs independently of the mass of the black hole (see footnote \ref{footnote:2} in Section \ref{sec:homologous}). A general relativistic extension of the model outlined here would provide a more accurate measure of the increase in the central density and temperature of the star for high-$\beta$ encounters between main sequence (and smaller) stars and black holes with masses $\gtrsim 10^{6}M_{\odot}$; the work of \citet{gafton19} shows that for the disruption of a solar-like, $\gamma = 5/3$ polytrope by a $10^6M_{\odot}$ supermassive black hole with $\beta = 11$ (for which $r_{\rm p} \simeq 4.29r_{\rm g}$, i.e., very near the marginally bound radius at which direct capture occurs), relativistic effects can increase the maximum-attained density of the compressing star by a factor of at most $\sim 2$ (i.e., depending on the spin of the black hole; though we note that these authors used a Newtonian potential for the stellar self-gravity, and relativistic self-gravity may modify these results at sufficiently close pericenters). We defer a relativistic extension of the model described here to a future investigation.}

\begin{acknowledgements}
E.R.C.~acknowledges support from the National Science Foundation through grant AST-2006684. C.J.N acknowledges funding from the European Union’s Horizon 2020 research and innovation program under the Marie Sk\l{}odowska-Curie grant agreement No 823823 (Dustbusters RISE project). Some of this research used the ALICE High Performance Computing Facility at the University of Leicester. Some of this work was performed using the DiRAC Data Intensive service at Leicester, operated by the University of Leicester IT Services, which forms part of the STFC DiRAC HPC Facility (\url{www.dirac.ac.uk}). The equipment was funded by BEIS capital funding via STFC capital grants ST/K000373/1 and ST/R002363/1 and STFC DiRAC Operations grant ST/R001014/1. DiRAC is part of the National e-Infrastructure. We used {\sc splash} \citep{Price:2007aa} for Figure \ref{fig:beta16_sim}. 
\end{acknowledgements}

\end{document}